\PassOptionsToPackage{pdfpagelabels=false}{hyperref}
\PassOptionsToPackage{plainpages=false}{hyperref}
\PassOptionsToPackage{draft}{hyperref}
\documentclass[a4paper,fleqn,usenatbib]{mnras}

\usepackage{newtxtext,newtxmath}

\usepackage[T1]{fontenc}
\usepackage{ae,aecompl}

\usepackage{graphicx}	
\usepackage{amsmath}	
\usepackage{amssymb}	

\usepackage{times}
\usepackage{longtable}
\usepackage{multirow}
\usepackage{rotating}
\usepackage[dvipsnames]{xcolor}
\usepackage{booktabs}
\usepackage{epstopdf}
\usepackage{bm}
\usepackage{indentfirst}
\usepackage{booktabs}
\usepackage{indentfirst}
\usepackage{threeparttable}
\usepackage{subfigure}
\subfiguretopcaptrue
\usepackage{mathrsfs}
\usepackage{url}

\title[Influence of atomic alignment on spectroscopy]{The influence of atomic alignment on absorption and emission spectroscopy}

\author[Zhang, Yan, Richter]{
Heshou Zhang,$^{1,2}$
Huirong Yan,$^{1,2}$\thanks{E-mail: hyan@mail.desy.de}
Philipp Richter,$^{2,3}$
\\
$^{1}$Deutsches Elektronen-Synchrotron DESY, Platanenallee 6, D-15738 Zeuthen, Germany\\
$^{2}$Institut f$\ddot{u}$r Physik und Astronomie, Universit$\ddot{a}$t Potsdam, Haus 28, Karl-Liebknecht-Str. 24/25, D-14476 Potsdam, Germany\\
$^{3}$Leibniz-Institut f$\ddot{u}$r Astrophysik Potsdam (AIP), An der Sternwarte 16, D-14482 Potsdam, Germany
}

\pubyear{2018}

\begin{document}
\label{firstpage}
\pagerange{\pageref{firstpage}--\pageref{lastpage}}
\maketitle

\begin{abstract}
Spectroscopic observations play essential roles in astrophysics. They are crucial for determining physical parameters in the universe, providing information about the chemistry of various astronomical environments. The proper execution of the spectroscopic analysis requires accounting for all the physical effects that are compatible to the signal-to-noise ratio. We find in this paper the influence on spectroscopy from the atomic/ground state alignment owing to anisotropic radiation and modulated by interstellar magnetic field, has significant impact on the study of interstellar gas. In different observational scenarios, we comprehensively demonstrate how atomic alignment influences the spectral analysis and provide the expressions for correcting the effect. The variations are even more pronounced for multiplets and line ratios. We show the variation of the deduced physical parameters caused by the atomic alignment effect, including alpha-to-iron ratio ([X/Fe]) and ionisation fraction. Synthetic observations are performed to illustrate the visibility of such effect with current facilities. A study of PDRs in $\rho$ Ophiuchi cloud is presented to demonstrate how to account for atomic alignment in practice. Our work has shown that due to its potential impact, atomic alignment has to be included in an accurate spectroscopic analysis of the interstellar gas with current observational capability.
\end{abstract}




\section{Introduction}

Spectroscopy plays a crucial role in studying the universe. Analysing atomic and ionic spectral line intensity\footnote{Throughout the paper, we address atoms and ions with the term "atoms" for the sake of simplicity.} is one of the most important part of spectroscopic study. Absorption and emission atomic lines have numerous applications in astronomy. They are the direct measurements of column density of lots of elements (e.g., \citealt{2013ApJ...772..110F,2013ApJ...772..111R,2016ARA&A..54..363D}), and thus help the analysis of chemical evolution and composition of astronomical structures. Physical parameters derived from the spectral line ratio (e.g., ionisation rate, temperature) facilitate us to understand the astrophysical environments better (see, e.g., \citealt{1978ApJS...36..595D,1995A&A...294..792H,2016ApJ...830..118S}). Progress has been made in modelling the physical processes, such as cloud extinction in Photodissociation regions (PDRs) and outflows of Active Galactic Nuclei (AGN), with atomic spectra (e.g., \citealt{1999ApJ...527..795K,2015ApJ...801...35C,2016AJ....151..173Y}).

The rapid development of spectroscopic observations requires an accurate measurement of the spectral lines that is compatible to the higher signal-to-noise (S/N) ratio. Therefore, all relevant physical effects that are capable to produce the spectral fluctuation larger than the noise amplitude have to be counted for the high precision observations. We will demonstrate in this paper the fluctuation of atomic/ionic lines from Interstellar Medium (ISM) induced by ground state alignment (GSA) effect can have significant impact on the physical parameter derivation. One may naively argue that the physical effect will be averaged out given the possible line-of-sight dispersion of the magnetic field line in the medium. Synthetic observations are performed accounting for the averaging and the results show that such argument does not hold. We shall demonstrate that the influence of GSA cannot be neglected for an accurate spectral line intensity analysis with current observational capability and provide analytical expressions for the correction.

This paper is organised as follows: The general physics for the modulation of spectral lines due to GSA is briefly discussed in \S 2. We then demonstrate in \S 3 the influence of radiative alignment on the spectrum, where the magnetic field does not exist/vary. In \S 4, we further discuss the impact of magnetic realignment, where the radiative geometry is fixed and the fluctuation of the spectral lines originates from the change of magnetic field. In addition, the influences of GSA on various astrophysical properties derived from spectral line ratios, including alpha-to-iron ratio ([X/Fe]) and ionisation fraction, are exhibited in \S 5. In order to illustrate how our modelling is applied to real observation, we apply in \S 6 our analytical tool to the processing of raw data from PDR in $\rho$ Ophiuchi cloud, showing how to remove the potential impact that GSA produces. Discussions are provided in \S 7 with the applicability of our modelling and conclusions are in \S 8.

\section{Atomic transitions under the influence of GSA}

\begin{figure}
\centering
\includegraphics[width=0.98\columnwidth]{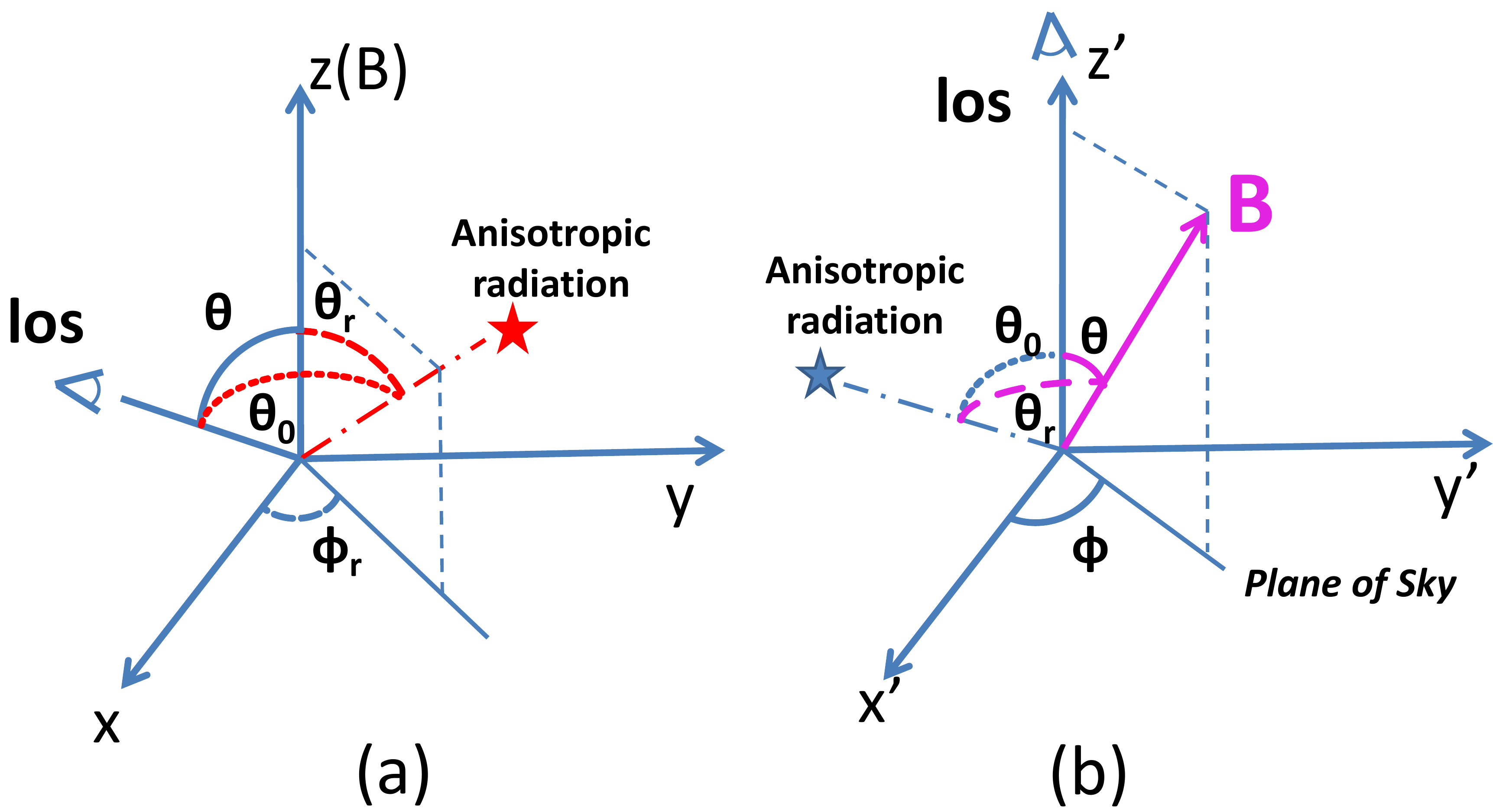}\label{scea}
\caption{(a) Theoretical frame $xyz-$coordinate system with the magnetic direction being $z-$axis and line of sight on $xz-$plane. The calculation is done in this frame. The direction of the anisotropic radiation is $\Omega_r\equiv(\theta_r,\phi_r)$, and that of the line of sight is $\Omega\equiv(\theta,\pi)$. (b) Observational frame $x'y'z'-$coordinate system with the line of sight being $z'-$axis, in which our results are presented. $x'y'-$plane is the plane of sky. $\theta_0$ is the angle between the direction of the incident radiation and the line of sight. The change of magnetic direction alters $\theta$ and $\theta_r$. }
\end{figure}

Substantial part of ISM are photon-excitation dominant areas, including PDRs, H\,{\sc ii} Region and reflection nebulae, etc. The spectroscopy from those areas will be influenced by GSA effect. The alignment here is in terms of the projection of the angular momenta of the atoms and it arises from the anisotropy in the radiation field. GSA happens when the optical/UV pumping transfers angular momentum from the radiation field to the atoms/ions.  The atomic angular momentum is along the incident radiation direction. Using point source as an example, the angular momentum of the photons that pump the atoms from the ground state to upper states can only be $\pm1$, whereas the isotropic spontaneous emission from upper states allows the angular momentum transfer $\pm1, 0$. Such difference leads to differential occupation among the magnetic sublevels on the ground state \citep{Landolfi:1986lh,YLfine,ZYD15}.

With the existence of magnetic field, which is ubiquitous in the universe, occupations among different sublevels are mixed according to the magnetic field direction due to the magnetic precession. The alignment is dependent on the comparison between magnetic precession rate ($2\pi\nu_L$) and photon excitation rate ($B\bar{J}^0_0$). For interstellar magnetic field, it is in GSA (saturated) regime with the magnetic precession prevailing over the photon excitation from the ground state ($2\pi\nu_L\gg B_{lu}\bar{J}^0_0$), but still too weak to influence the excited level. It has been demonstrated that for a wide range of magnetic field in ISM and IGM ($10^{-15}G \la B \la mG $), $A\gg2\pi\nu_L\gg B_{lu}\bar{J}^0_0$ and thus GSA applies \citep{YLhyf,YLHanle, 2013Ap&SS.343..335S, ZY2018subP}. We show in \S 7 that GSA is indeed the dominant regime by providing the division boundary between the GSA (saturated) regime and the ground-level Hanle regime, where magnetic field strength also matters.

We present below the expressions for the modulation of absorption and emission coefficients \footnote{Readers may directly go to \S 3 if they are only interested in observational consequences.}. The atomic density of different angular momentum is depicted by the density matrix tensor, e.g., for the lower level $J_l$ it is denoted as $\rho^k_q(J_l)$\footnote{For example, the irreducible tensor for $J/F=1$ is $\rho_0^2=[\rho(1,1)-2\rho(1,0)+\rho(1,-1)]$ (see, e.g., \citealt{1957RvMP...29...74F,1965JETP...21..227D,1978A&A....69...57B}).}. Only unpolarized incoming light is considered and therefore $k=0, 2$ here. The ratio between the density tensor and the total density of the level is defined as the alignment parameter $\sigma^k_q\equiv\rho^k_q/\rho^0_0$. In the regime of GSA, $q$ is always $0$ for the levels on the ground state. It is most convenient to calculate the GSA in the theoretical frame $xyz-$system (Fig.1a), where the magnetic direction is $z-$axis. The direction of the line of sight is denoted as $\Omega$ and $\mathcal{J}^K_Q(i,\Omega)$ is the irreducible geometric tensor for the observed light.
Due to GSA, the absorption coefficients of the atomic transitions are expressed by (\citealt{Landi-DeglInnocenti:1984kl}, see also \citealt{YLfine}):
\begin{equation}\label{etai}
\eta_i(\nu,\Omega)=\frac{h\nu_0}{4\pi}B_{lu}n(J_l)\Psi(\nu-\nu_0)\sum_{\substack{K}}(-1)^K\omega^{K}_{J_lJ_u}\sigma^K_0(J_l)\mathcal{J}^K_0(i,\Omega),
\end{equation}
where $K=0,2$, $\omega^0_{J_1J_2}\equiv1$, $\omega^2_{J_1J_2}\equiv\{1,1,2;J_1,J_1,J_2\}/\{1,1,0;J_1,J_1,J_2\}$, in which the matrix with braces "$\{\}$" represent the $6-j$ symbol. The quantity $B_{lu}$ is the Einstein coefficient for absorption\footnote{The data of Einstein coefficients used in the paper are taken from the Atomic Line List (\url{http://www.pa.uky.edu/~peter/atomic/}) and the NIST Atomic Spectra Database.}. The total atomic population $n(J_l)$ on the lower level $J_l$ is defined as $n\sqrt{[J_l]}\rho^0_0(J_l)$, where $[j]=2j+1$. $\Psi(\nu-\nu_0)$ is the line profile. As shown in Eq.~\eqref{etai}, the radiative pumping and the magnetic direction will modulate the absorption coefficients $\eta_i$ and thus the absorption spectrum varies.

For resonance emission lines, the excited states are influenced by the differential occupation on the ground state through radiative excitation (see, e.g., \citealt{YLHanle}). In diffuse ISM and IGM, the magnetic field is weak and the decay rate of atoms from the levels on the excited states is much higher than the magnetic precession rate ($A\gg2\pi\nu_L$). Thus, $q$ can be non-zero for the density tensor of the atoms on the excited states. the emission coefficients $\epsilon_i(i=0\sim3)$ from the upper level $J_u$ are \citep[see][]{YLhyf}:
\begin{equation}\label{epsiloni}
\epsilon_i(\nu,\Omega)=\frac{h\nu_0}{4\pi}A_{ul}n(J_u)\Psi(\nu-\nu_0)\sum_{\substack{KQ}}\omega^{K}_{J_uJ_l}\sigma^K_Q(J_u)\mathcal{J}^K_Q(i,\Omega),
\end{equation}
where $K=0,2;Q=0,\pm1,\pm2$. The quantity $A_{ul}$ is the Einstein coefficient for emission. Similar to the case of absorption, the emissivity $\epsilon_0$ is also influenced by the anisotropic radiative pumping and magnetic field, as demonstrated in Eq.~\eqref{epsiloni}. In the following, we will evaluate quantitatively the influence of GSA on absorption and emission spectral lines in the observational frame (see Fig.1b).

\section{Influence of radiative alignment on the spectrum intensity}

Spectral line intensity are modulated due to the GSA induced by the anisotropic radiation. We first consider the unmagnetized case, and study only the dependence modulation on the scattering angle $\theta_0$, the angle between line of sight and the incident radiation direction. It is worth noting that there is no alignment (i.e., the alignment parameter $\sigma^2_0=0$) at $\theta_0=54.7^{\circ}$ or $180^\circ-54.7^\circ$ (Van Vleck angle, \citealt{1925PNAS...11..612V,1974PASP...86..490H}), the corresponding intensity $I_{VV}$ is used as the ``standard'' for comparison in this section.

\begin{figure*}
\centering
\subfigure[]{
\includegraphics[width=0.48\textwidth]{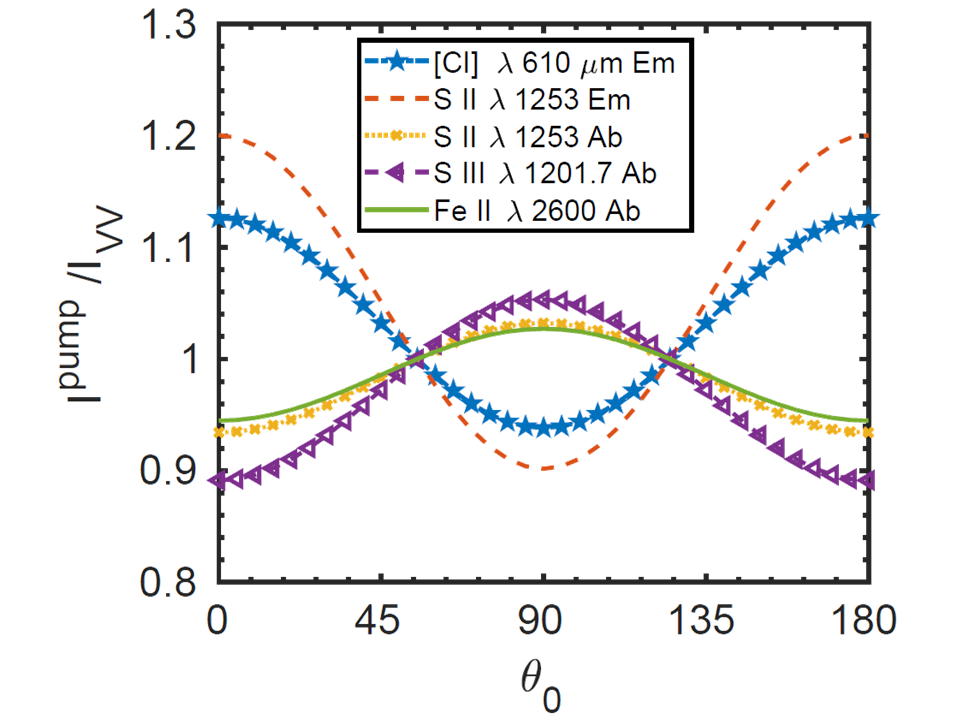}\label{purpump}}
 \subfigure[]{
\includegraphics[width=0.48\textwidth]{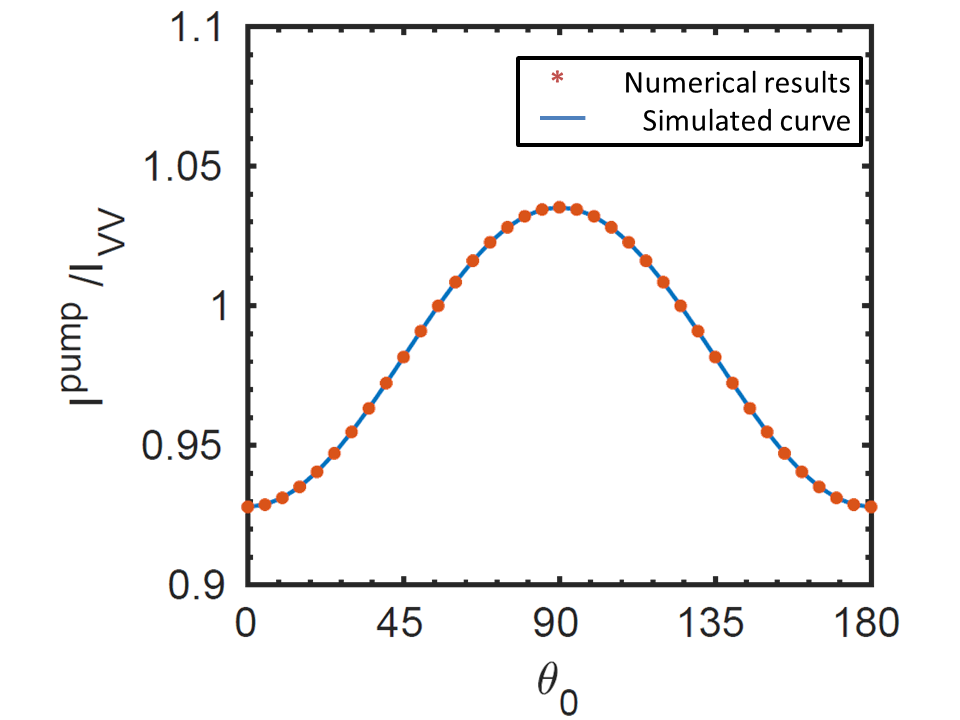}\label{pumpfitplot}}
\caption{(a) The modulation of several selected atomic lines owing to radiative alignment; (b) The fitting of the analytical expression for Fe\,{\sc ii}$\lambda2600\mbox{\AA}$ emission line.
}\label{ducloud}
\end{figure*}

The intensity fluctuations due to radiative alignment for several selected spectral lines are plotted in Fig.~\ref{purpump}. We find that the intensity fluctuation ratio due to radiative alignment can be fitted with a simple analytical expression with $5\sigma$ confidence:
\begin{equation}\label{ratiopurpump}
r^{pump}(\theta_0)\equiv I^{pump}/I_{VV}=p_1+p_2\cos2\theta_0,
\end{equation}
where $p_1, p_2$ vary with spectral lines. The fitting for Fe\,{\sc ii}$\lambda2600\mbox{\AA}$ is presented as an example in Fig.~\ref{pumpfitplot}. The comprehensive fitting parameters for various transitions are presented in Table~\ref{pumpfit}.

\begin{table}
\centering
\caption{FITTING PARAMETERS FOR RADIATIVE ALIGNMENT}\label{pumpfit}
\begin{tabular}{|c|c|c|c|c|c|}
\hline
\hline
Species & $\lambda$ & \multicolumn{2}{|c|}{Absorption} & \multicolumn{2}{|c|}{Emission} \\
        & $(\mbox{\AA})$ & $p^{ab}_1$ & $p^{ab}_2$ & $p^{em}_1$ & $p^{em}_2$ \\
\hline
\hline
C\,{\sc i} & $1193.03$ & 1 & 0 & 1.0307 & 0.0898 \\
\hline
C\,{\sc ii} & $1334.53$ & 0 & 0 & 1.0451 & 0.1320 \\
\hline
O\,{\sc i} & $1025.76$ & 0.9984 & -0.0047 & 1.0194 & 0.0566 \\
\hline
\multirow{3}{*}{S\,{\sc i}} & $1474.57$ & 1.0376 & 0.1099 & 1.0032 & 0.0093 \\
\cline{2-6}
& $1474.38$ & 0.9625 & -0.1097 & 0.9776 & -0.0655 \\
\cline{2-6}
& $1473.99$ & 1.0107 & 0.0314 & 1.0350 & 0.1024 \\
\hline
\multirow{3}{*}{S\,{\sc ii}} & $1250.58$ & 1.0210 & 0.0613 & 1 & 0 \\
\cline{2-6}
& $1253.81$ & 0.9832 & -0.0490 & 1.0510 & 0.1494 \\
\cline{2-6}
& $1259.52$ & 1.0042 & 0.0123 & 1.0378 & 0.1104 \\
\hline
Si\,{\sc ii} & $1190.42$ & 1 & 0 & 1.0440 & 0.1286 \\
\hline
S\,{\sc iii} & $1012.50$ & 1 & 0 & 1.0265 & 0.0774 \\
\hline
S\,{\sc iv} & $1062.66$ & 1 & 0 & 1.0623 & 0.1823 \\
\hline
\multirow{3}{*}{Fe\,{\sc ii}} & $1142.36$ & 1.0097 & 0.0283 & 0.9816 & -0.0537 \\
\cline{2-6}
& $1143.22$ & 0.9859 & -0.0411 & 0.9817 & -0.0535 \\
\cline{2-6}
& $1144.93$ & 1.0053 & 0.0154 & 1.0230 & 0.0673 \\
\hline
\hline
Species & $\lambda$ & & \multicolumn{2}{|c|}{Emission} & \\
        & $({\mu}m)$ & & $p^{em}_1$ & $p^{em}_2$ & \\
\hline
\hline
[C\,{\sc i}] & $610$ & & 1.0332 & 0.0940 & \\
\hline
[C\,{\sc ii}] & $157.7$ & & 0.9795 & -0.0599 & \\
\hline
[O\,{\sc i}] & $63.2$ & & 1.0046 & 0.0135 & \\
\hline
[Si\,{\sc ii}] & $34.8$ & & 1.0206 & 0.0602 & \\
\hline
[S\,{\sc i}] & $25.2$ & & 1.0030 & 0.0089 & \\
\hline
[S\,{\sc iii}] & $33.5$ & & 1.0325 & 0.0951 & \\
\hline
[S\,{\sc iv}] & $10.5$ & & 1.0061 & 0.0177 & \\
\hline
[Fe\,{\sc ii}] & $26.0$ & & 1.0049 & 0.0143 & \\
\hline
\hline
\end{tabular}
\end{table}

\begin{figure*}
\begin{center}
\begin{tabular}{ll}
\begin{tabular}{l}
  \subfigure[]{
\includegraphics[width=0.27\textwidth]{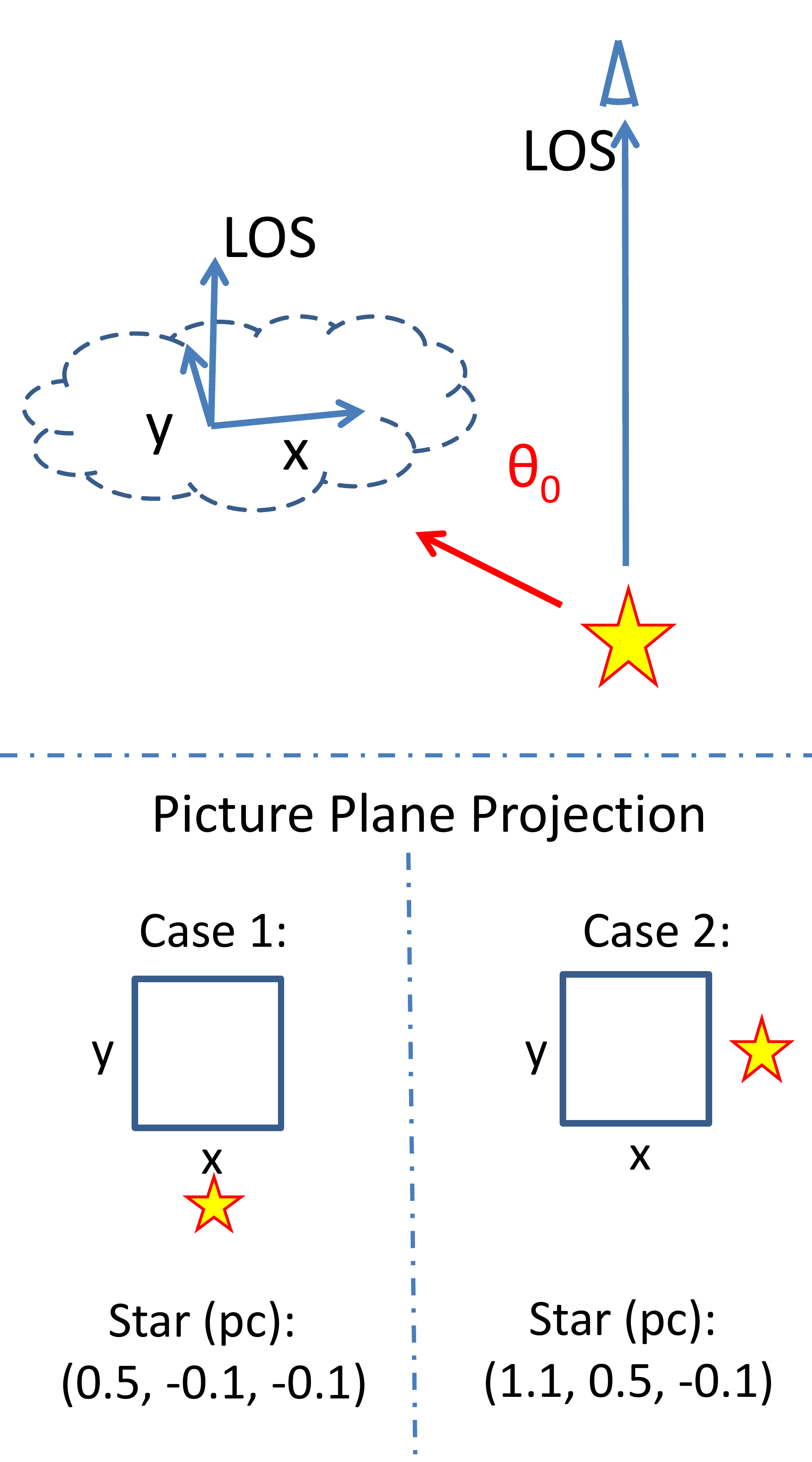}\label{radigeo}}
\end{tabular}
\begin{tabular}{ll}
 \subfigure[Case 1]{
\includegraphics[width=0.31\textwidth]{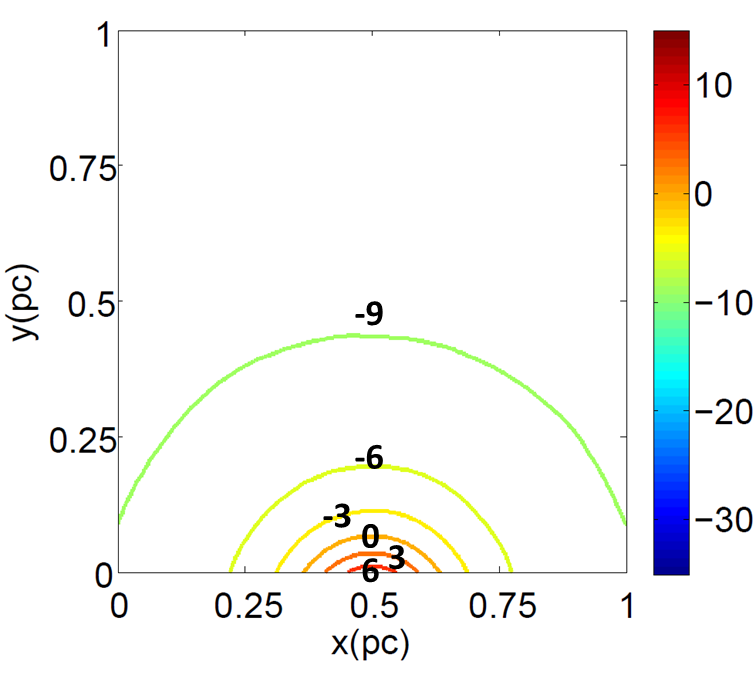}\label{radicase1}}&
\subfigure[Case 2]{
\includegraphics[width=0.31\textwidth]{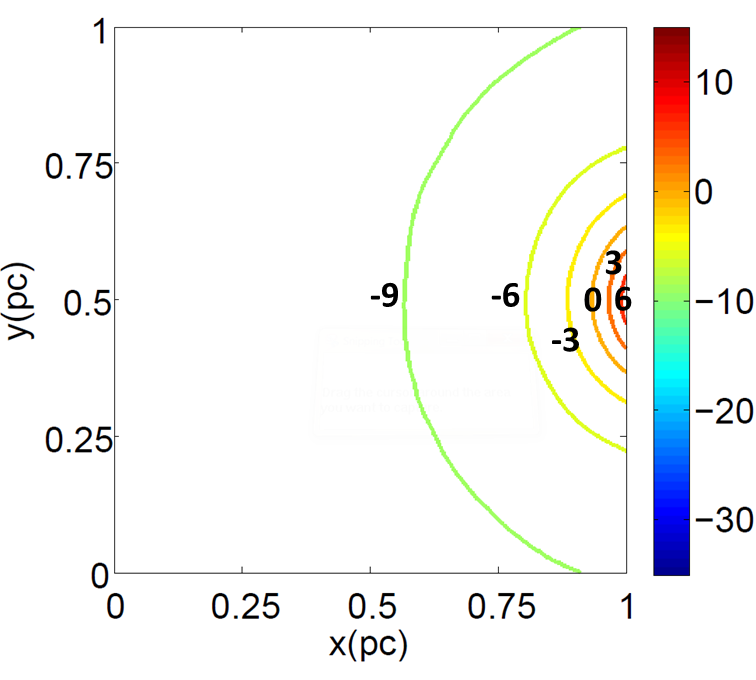}\label{radicase2}}\\
 \subfigure[Case 1 with magnetic field]{
\includegraphics[width=0.31\textwidth]{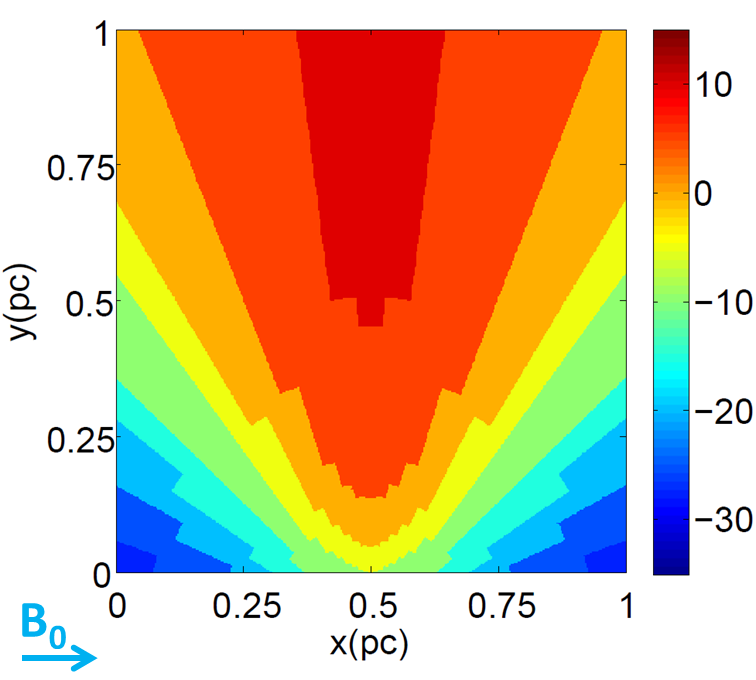}\label{radiBcase1}}&
\subfigure[Case 2 with magnetic field]{
\includegraphics[width=0.31\textwidth]{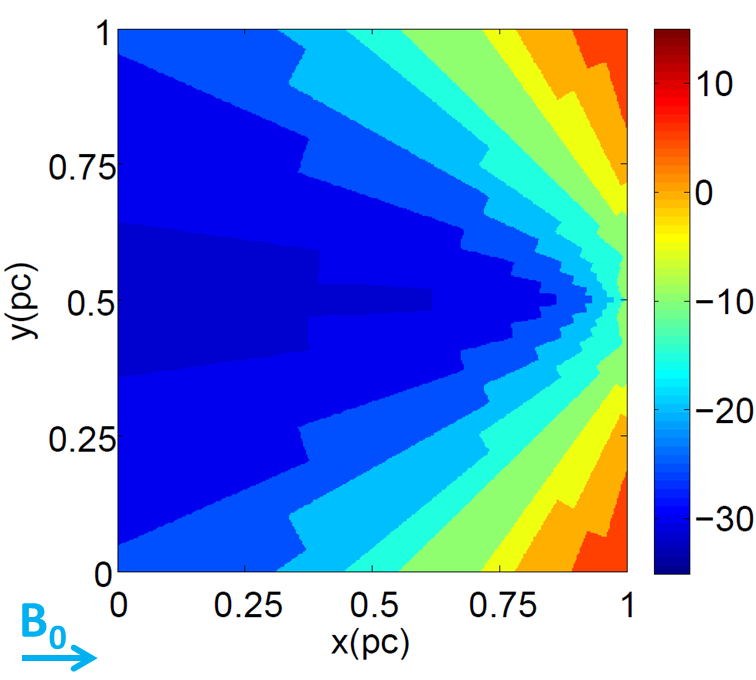}\label{radiBcase2}}\\
\end{tabular}
\end{tabular}
\end{center}
\caption{The influence of radiative alignment. Magnetic field is not accounted for. (a) The geometric illustration, the synthetic ISM is illuminated by the O-type star from different positions, the picture plane projection is on the right; (b,c) variations of the line ratio of two components in the S\,{\sc i} triplets $\lambda\sim1474\mbox{\AA}$ ($1473.99, 1474.38\mbox{\AA}$) for case 1,2, respectively; (d,e) same as (b,c), but with constant magnetic field along $x-$axis included.}
\end{figure*}

The influence is more significant for multiplets and line ratios since it varies among different lines. We perform synthetic observation on a cloud that is illuminated by a star located at two different positions ({\it case 1} and {\it case 2}, whose picture plane projections are marked in Fig.~\ref{radigeo}). The resulting fluctuations of the line ratio of two components in the S\,{\sc i} triplets $\lambda\sim1474\mbox{\AA}$ are shown in Fig.~\ref{radicase1} and Fig.~\ref{radicase2}, respectively. Moreover, the possible fluctuations will be further enhanced if magnetic field exists. Fig.~\ref{radiBcase1} and Fig.~\ref{radiBcase2} are the results of synthetic observations for case 1 and 2 with the same magnetic field. The fluctuation of the line ratio ranges from overestimation of $\gtrsim30\%$ to underestimation of $\gtrsim10\%$. By comparing Fig.~\ref{radiBcase1} and Fig.\ref{radiBcase2}, we further conclude that the influence of GSA on the same spectral line also varies according to radiation geometry even with the same underlying magnetic field.

We define $R_{th}$ to depict the possible intensity fluctuation range introduced by GSA:
\begin{equation}\label{allethre}
R_{th}\equiv \frac{max\{(I_{VV}+\Delta I)\}}{min\{(I_{VV}+\Delta I)\}},
\end{equation}
The comprehensive results for the $R_{th}$ are shown in Table~\ref{allRth}. Owing to GSA, the same column density could produce an intensity variation by a factor of $\gtrsim 1.5$, which is far beyond the noise amplitude of current high S/N telescope.

\begin{table}
\centering
\caption{THE RANGE OF FLUCTUATION ARISING FROM GSA}\label{allRth}
\begin{tabular}{|c|c|c|c|}
\hline
\hline
Species & $\lambda$ & {Absorption} & {Emission} \\
        & $(\mbox{\AA})$ & $R^{ab}_{th}$ & $R^{em}_{th}$  \\
\hline
\hline
C\,{\sc i} & $1193.03$ & 1 & 1.3004 \\
\hline
C\,{\sc ii} & $1334.53$ & 1 & 1.4275 \\
\hline
O\,{\sc i} & $1025.76$ & 1.0682 & 1.1834 \\
\hline
\multirow{3}{*}{S\,{\sc i}} & $1474.57$ & 1.4121 & 1.0677 \\
\cline{2-4}
& $1474.38$ & 1.3264 & 1.2442 \\
\cline{2-4}
& $1473.99$ & 1.2526 & 1.6439 \\
\hline
\multirow{3}{*}{S\,{\sc ii}} & $1250.58$ & 1.1278 & 1 \\
\cline{2-4}
& $1253.81$ & 1.1049 & 1.3949 \\
\cline{2-4}
& $1259.52$ & 1.0247 & 1.3234 \\
\hline
Si\,{\sc ii} & $1190.42$ & 1 & 1.4261 \\
\hline
S\,{\sc iii} & $1012.50$ & 1 & 1.4726 \\
\hline
S\,{\sc iv} & $1062.66$ & 1 & 1.5979 \\
\hline
\multirow{3}{*}{Fe\,{\sc ii}} & $1142.36$ & 1.1560 & 1.1216 \\
\cline{2-4}
& $1143.22$ & 1.1054 & 1.3509 \\
\cline{2-4}
& $1144.93$ & 1.0361 & 1.1904 \\
\hline
\hline
Species & $\lambda$ & \multicolumn{2}{|c|}{Emission} \\
        & $({\mu}m)$ & \multicolumn{2}{|c|}{$R^{em}_{th} (\%)$}  \\
\hline
\hline
[C\,{\sc i}] & $610$ & \multicolumn{2}{|c|}{1.3632} \\
\hline
[C\,{\sc ii}] & $157.7$ & \multicolumn{2}{|c|}{1.1307} \\
\hline
[O\,{\sc i}] & $63.2$ & \multicolumn{2}{|c|}{1.1217} \\
\hline
[Si\,{\sc ii}] & $34.8$ & \multicolumn{2}{|c|}{1.1338} \\
\hline
[S\,{\sc i}] & $25.2$ & \multicolumn{2}{|c|}{1.1464} \\
\hline
[S\,{\sc iii}] & $33.5$ & \multicolumn{2}{|c|}{1.3712} \\
\hline
[S\,{\sc iv}] & $10.5$ & \multicolumn{2}{|c|}{1.0358} \\
\hline
[Fe\,{\sc ii}] & $26.0$ & \multicolumn{2}{|c|}{1.0344} \\
\hline
\hline
\end{tabular}
\end{table}

\section{Influence of magnetic realignment on spectral lines}

In this section we investigate the fluctuation of spectral lines from the magnetic realignment. The variation ratio of the line intensity $r_{mag}=I_{ob}/I_{0}$ is used to characterize the influence of magnetic field, where $I_{ob}$ is the actual line intensity observed and $I_{0}$ corresponds to the supposed line intensity without the magnetic realignment (or $\theta_r=0$).

\begin{figure*}
\centering
 \subfigure[$r^{ab}(1250\mbox{\AA})$ variation, $\theta_0=90^{\circ}$]{
\includegraphics[width=0.64\columnwidth,
 height=0.24\textheight]{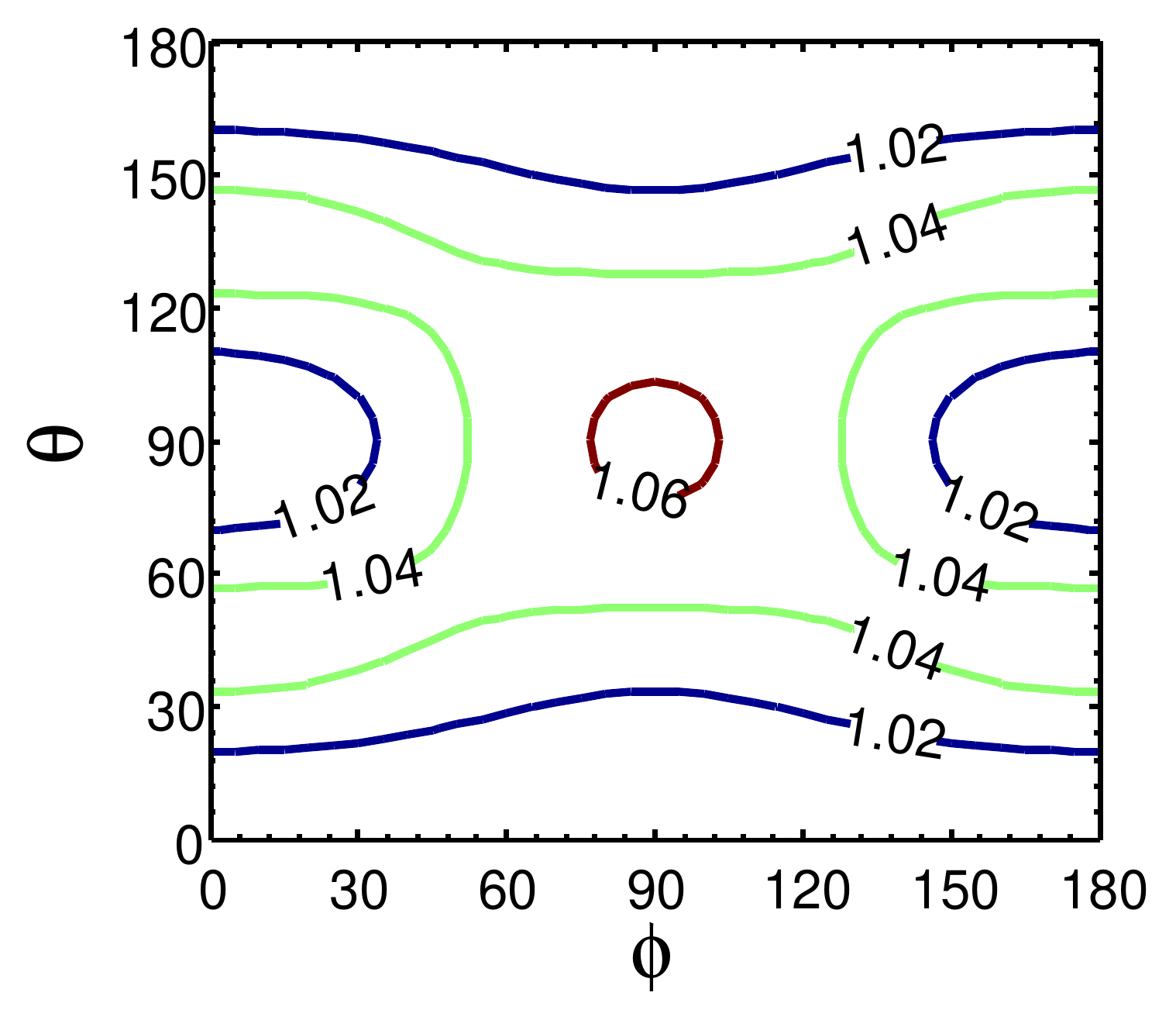}\label{S2bbrabv1c90}}
\subfigure[$r^{em}(1143\mbox{\AA})$ variation, $\theta_0=90^{\circ}$]{
\includegraphics[width=0.64\columnwidth,
 height=0.24\textheight]{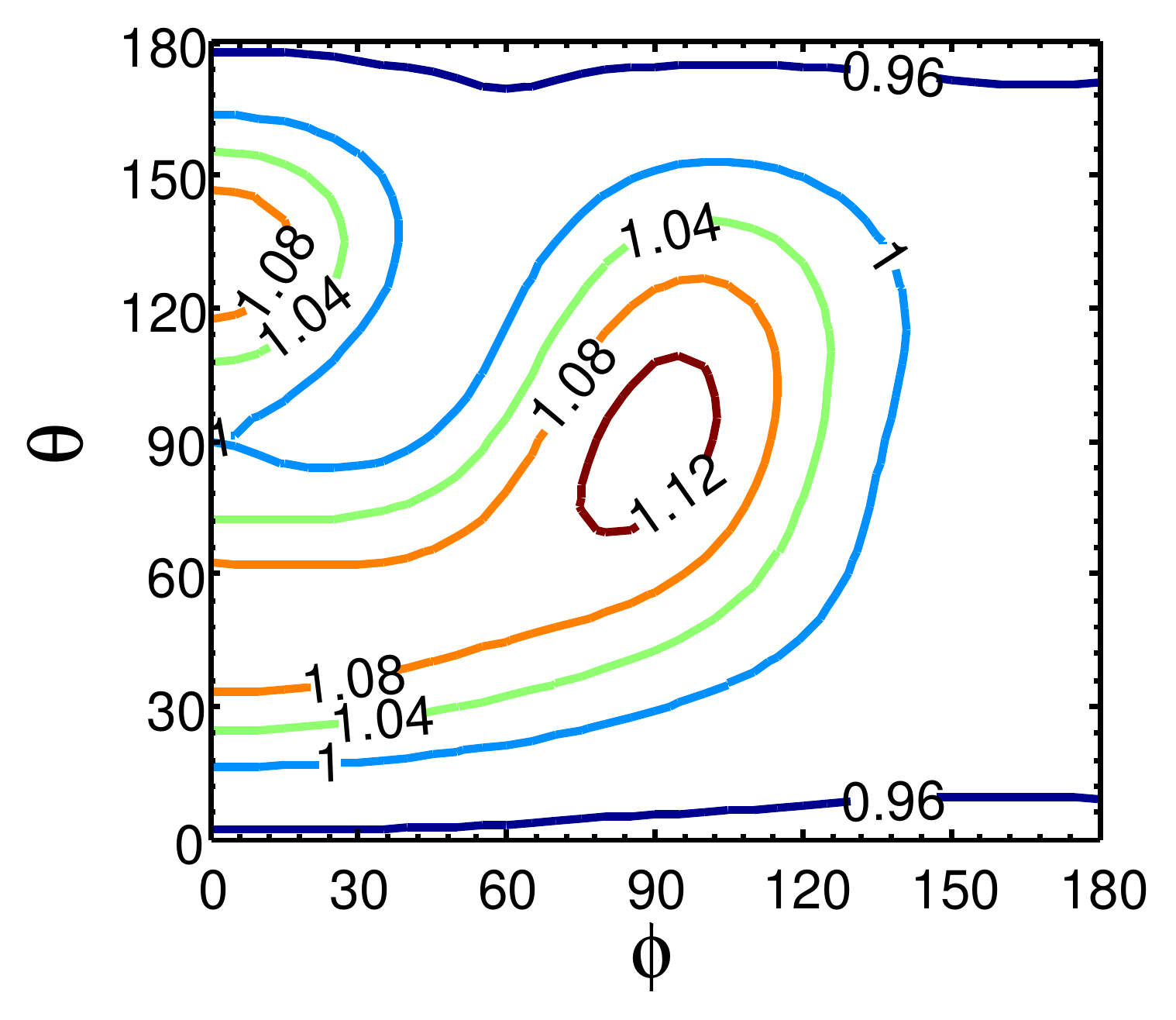}\label{Fe2bbremv11to9y6fc90}}
 \subfigure[$r^{em}(610{\mu}m)$ variation, $\theta_0=90^{\circ}$]{
\includegraphics[width=0.64\columnwidth,
 height=0.24\textheight]{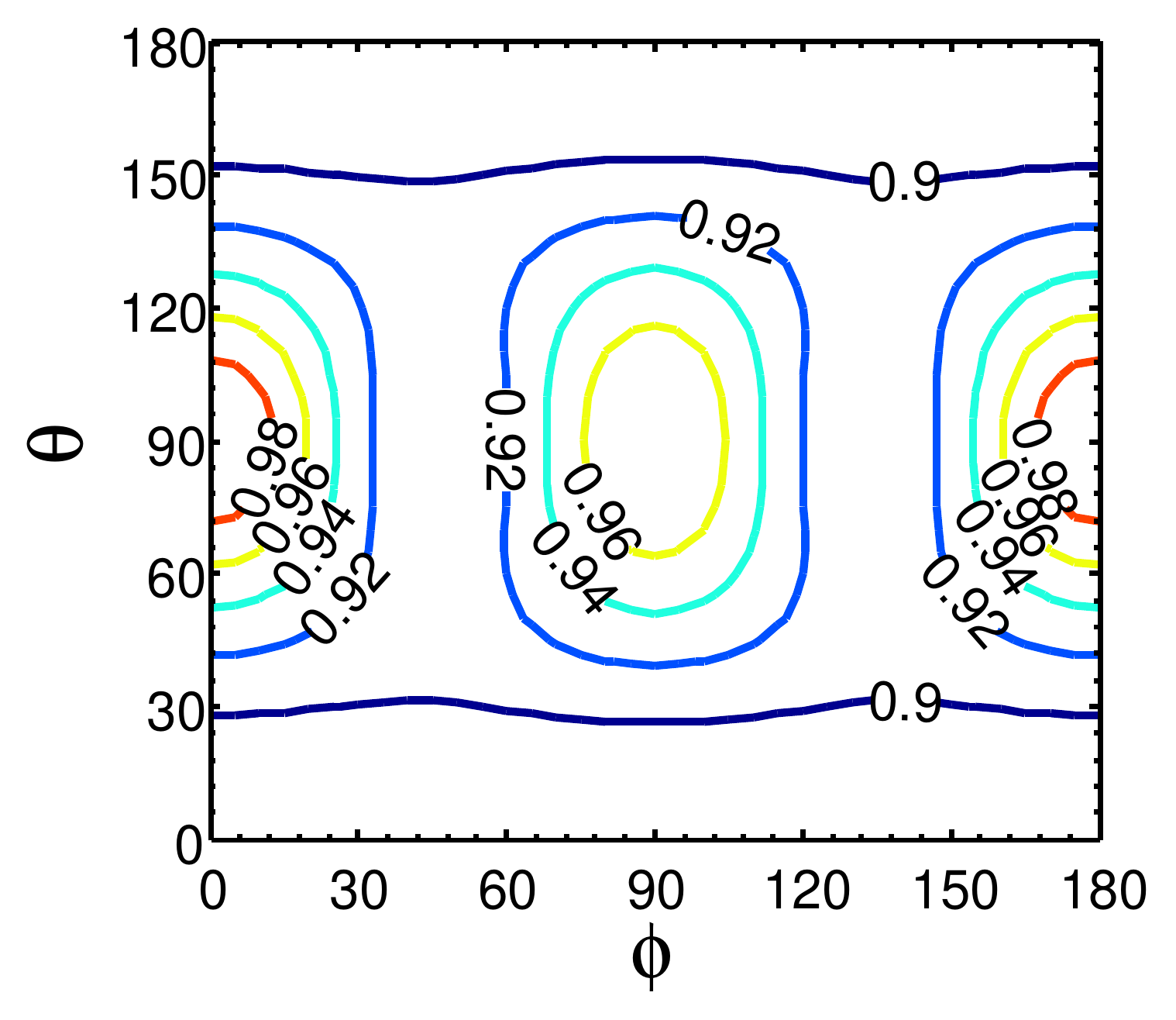}\label{C1bbrsubem1c90}}
\caption{The variation ratio of the line intensity under the influence of magnetic fields for (a) S\,{\sc ii}$\lambda1250\mbox{\AA}$ absorption, (b) Fe\,{\sc ii}$\lambda1143\mbox{\AA}$ emission, and (c) C\,{\sc i} $\lambda610{\mu}m$ emission, respectively. The line of sight is vertical to the direction of the incident radiation ($\theta_0=90^{\circ}$), which corresponds to a face-on disk.}\label{uvcp}
\end{figure*}

\subsection{UV/optical spectra}
\label{Allowed_tr}

\begin{figure}
\centering
\includegraphics[width=0.99\columnwidth]{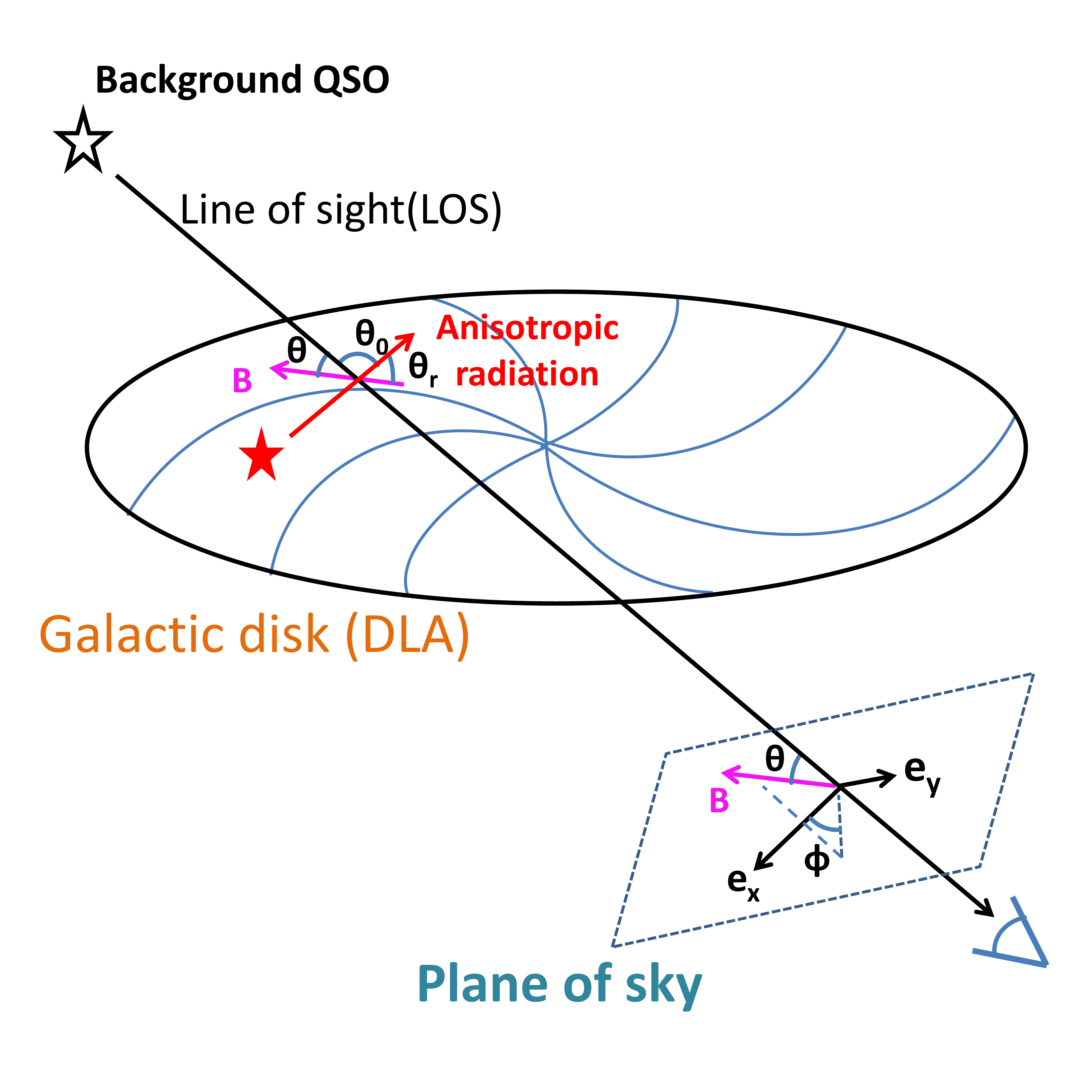}
\caption{Schematics for the typical environment where atomic lines are altered by magnetic fields. Presented is a DLA region on the spiral arm of a galaxy.}\label{scena}
\end{figure}

As an example, we show in Fig.\ref{scena} a typical scenario, in which the effect of magnetic realignment leads to the measurable variation in the observed line intensity. Consider a typical late-type galaxy with an extended neutral gas disk and an interstellar magnetic field. Star-formation is expected to take place in distinct regions in the disk (i.e., in spiral arms). Therefore, the interstellar radiation field is likely to be anisotropic at a given point in the disk. In the spectrum of the background point source (e.g., a Quasi-Stellar Object, hereafter QSO), the gas disk will leave its imprint through many absorption lines from neutral and ionised species, where most of the lines are located in the (rest frame) UV. Such disk absorbers are known to contribute to the population of the so-called DLA absorbers that are frequently observed at low and high redshift in QSO spectra. The DLA absorbers adopted to illustrate the influence of magetic fields on the spectroscopy are dominated by cold neutral hydrogen \citep[see, e.g.,][]{2000ApJS..130....1R}. As long as the angle between the ambient magnetic field and the direction of the incident radiation at the place where the sightline pierces the disk is not Van Vleck angle, the GSA will cause changes in the central absorption depths of unsaturated absorption lines. The variation of the line intensity of S\,{\sc ii}$\lambda1250.58\mbox{\AA}$ with respect to different magnetic field direction is presented in Fig.\ref{S2bbrabv1c90} given the line of sight perpendicular to the direction of the incident radiation ($\theta_0=90^\circ$).

As shown in Fig.\ref{S2bbrabv1c90}, {\it the spectra observed change with the direction of magnetic field.} In addition, the error of the observation for S\,{\sc ii}$\lambda1250.58\mbox{\AA}$ due to photon noise is $2.5\%$ in current observation (see \citealt{2007ApJS..171...29P}), which is smaller than the variations due to GSA in most of the areas- top at $7\%$- in Fig.\ref{S2bbrabv1c90}. The maximum enhancement and reduction caused by magnetic realignment for different absorption lines are presented in Table \ref{absorpretable}, in which the absorptions are from all the levels on the ground state that have much longer life time than magnetic precession period\footnote{These lines are detected at the places such as galactic halos \citep{2013ApJ...772..111R,2014ApJ...787..147F} and circumstellar medium in GRB\citep{2006A&A...451L..47F}.}. The fluctuation of intensity $\Delta\rm{I}/\rm{I_0}=r^{ab}_{mag}-1$. The influence of magnetic fields are different among the lines. For some of the lines, the variation is more than $20\%$. Thus, The measurable changes of absorption spectrum can be introduced due to magnetic realignment. The influence of GSA on the derived alpha-to-iron ratio of the DLA absorbers will be discussed in \S 5 with real observation analysis.

\begin{table}
\centering
\caption{VARIATIONS OF UV/OPTICAL ABSORPTION LINES DUE TO GSA}\label{absorpretable}
\begin{tabular}{|c|c|c|c|c|c|}
\hline
\multicolumn{6}{|c|}{Absorption from the ground level of the ground state}\\
\hline
\hline
Species & $\lambda$ & $(\Delta\rm{I}/\rm{I_0})_{min}$ & $\theta_{0min}$ & $(\Delta\rm{I}/\rm{I_0})_{max}$ & $\theta_{0max}$ \\
 &  $(\mbox{\AA})$ & $\%$ & & $\%$ &  \\
\hline
O\,{\sc i} & $1025.76$ & $-6.37$ & $90^{\circ}$ & $+0$ & $0^{\circ}$\\
\hline
\multirow{3}{*}{S\,{\sc i}} & $1474.57$ & $-27.29$ & $0^{\circ}$ & $+1.81$ & $54.7^{\circ}$\\
\cline{2-6}
& $1474.38$ & $-22.82$ & $90^{\circ}$ & $+2.62$ & $30^{\circ}$\\
\cline{2-6}
& $1473.99$ & $-19.99$ & $0^{\circ}$ & $+0.08$ & $54.7^{\circ}$\\
\hline
\multirow{3}{*}{S\,{\sc ii}} & $1250.58$ & $-7.56$ & $0^{\circ}$ & $+6.33$ & $90^{\circ}$\\
\cline{2-6}
& $1253.81$ & $-4.71$ & $90^{\circ}$ & $+7.00$ & $0^{\circ}$\\
\cline{2-6}
& $1259.52$ & $-1.61$ & $0^{\circ}$ & $+1.22$ & $90^{\circ}$\\
\hline
\multirow{3}{*}{Ti\,{\sc ii}} & $3072.97$ & $-1.64$ & $0^{\circ}$ & $+1.30$ & $90^{\circ}$\\
\cline{2-6}
& $3241.98$ & $-1.03$ & $90^{\circ}$ & $+1.35$ & $0^{\circ}$\\
\cline{2-6}
& $3229.19$ & $-0.33$ & $0^{\circ}$ & $+0.26$ & $90^{\circ}$\\
\hline
\multirow{3}{*}{Fe\,{\sc ii}} & $1142.36$ & $-1.69$ & $0^{\circ}$ & $+4.42$ & $90^{\circ}$\\
\cline{2-6}
& $1143.22$ & $-2.86$ & $90^{\circ}$ & $+7.94$ & $0^{\circ}$\\
\cline{2-6}
& $1144.93$ & $-0.46$ & $35.3^{\circ}$ & $+3.21$ & $90^{\circ}$\\
\hline
\hline
\multicolumn{6}{|c|}{Absorption from metastable levels of the ground state}\\
\hline
\hline
\multirow{3}{*}{C\,{\sc i}$^\ast$} & $1260.93$ & $-24.20$ & $0^{\circ}$ & $+1.54$ & $54.7^{\circ}$\\
\cline{2-6}
& $1261.00$ & $-17.45$ & $90^{\circ}$ & $+0.46$ & $35.3^{\circ}$\\
\cline{2-6}
& $1261.12$ & $-15.76$ & $0^{\circ}$ & $+0$ & $0^{\circ}$\\
\hline
\multirow{3}{*}{C\,{\sc ii}$^\ast$} & $1037.02$ & $-9.02$ & $90^{\circ}$ & $+3.68$ & $0^{\circ}$\\
\cline{2-6}
& $1335.66$ & $-10.34$ & $0^{\circ}$ & $+1.29$ & $54.7^{\circ}$\\
\cline{2-6}
& $1335.71$ & $-5.39$ & $90^{\circ}$ & $+0.10$ & $54.7^{\circ}$\\
\hline
O\,{\sc i}$^\ast$ & $1027.43$ & $-9.57$ & $90^{\circ}$ & $+0$ & $0^{\circ}$\\
\hline
\multirow{2}{*}{Si\,{\sc ii}$^\ast$} & $1264.74$ & $-5.59$ & $90^{\circ}$ & $+5.32$ & $0^{\circ}$\\
\cline{2-6}
& $1265.00$ & $-3.00$ & $0^{\circ}$ & $+0.27$ & $54.7^{\circ}$\\
\hline
\multirow{3}{*}{S\,{\sc i}$^\ast$} & $1303.11$ & $-21.05$ & $0^{\circ}$ & $+1.72$ & $54.7^{\circ}$\\
\cline{2-6}
& $1302.86$ & $-14.43$ & $90^{\circ}$ & $+0.50$ & $35.3^{\circ}$\\
\cline{2-6}
& $1302.34$ & $-12.74$ & $0^{\circ}$ & $+0$ & $0^{\circ}$\\
\hline
\multirow{3}{*}{S\,{\sc iii}$^\ast$} & $1015.50$ & $-24.67$ & $0^{\circ}$ & $+1.53$ & $54.7^{\circ}$\\
\cline{2-6}
& $1015.57$ & $-17.88$ & $90^{\circ}$ & $+0.46$ & $54.7^{\circ}$\\
\cline{2-6}
& $1015.78$ & $-16.19$ & $0^{\circ}$ & $+0$ & $0^{\circ}$\\
\hline
\multirow{2}{*}{S\,{\sc iv}$^\ast$} & $1072.97$ & $-1.20$ & $90^{\circ}$ & $+2.20$ & $0^{\circ}$\\
\cline{2-6}
& $1073.52$ & $-0.20$ & $90^{\circ}$ & $+0.56$ & $90^{\circ}$\\
\hline
\multirow{3}{*}{C\,{\sc i}$^{\ast\ast}$} & $1193.65$ & $-24.65$ & $0^{\circ}$ & $+1.06$ & $54.7^{\circ}$\\
\cline{2-6}
& $1193.39$ & $-20.93$ & $90^{\circ}$ & $+1.51$ & $35.3^{\circ}$\\
\cline{2-6}
& $1193.24$ & $-18.74$ & $0^{\circ}$ & $+0$ & $0^{\circ}$\\
\hline
\multirow{3}{*}{S\,{\sc iii}$^{\ast\ast}$} & $1202.12$ & $-24.20$ & $0^{\circ}$ & $+0.95$ & $54.7^{\circ}$ \\
\cline{2-6}
& $1201.73$ & $-20.62$ & $90^{\circ}$ & $+1.37$ & $35.3^{\circ}$\\
\cline{2-6}
& $1200.97$ & $-18.53$ & $0^{\circ}$ & $+0$ & $0^{\circ}$\\
\hline
\end{tabular}
\begin{tablenotes}
      \small
      \item Note: The comparison is made by first considering a specific $\theta_0$ and changing the magnetic direction in the full space to find the maximum enhancement and reduction for the chosen $\theta_0$. Then we scan $\theta_0$ for all the possible angles and compare the maximum variation for different $\theta_0$. $(\Delta \rm{I}/\rm{I_0})_{max}$ means the maximum enhancement of the observed intensity due to magnetic realignment and $(\Delta \rm{I}/\rm{I_0})_{min}$ means the reduction. The corresponding geometry are denoted as $\theta_{0max}$ and $\theta_{0min}$, respectively.
\end{tablenotes}
\end{table}

\begin{figure*}
\centering
\includegraphics[width=1.8\columnwidth]{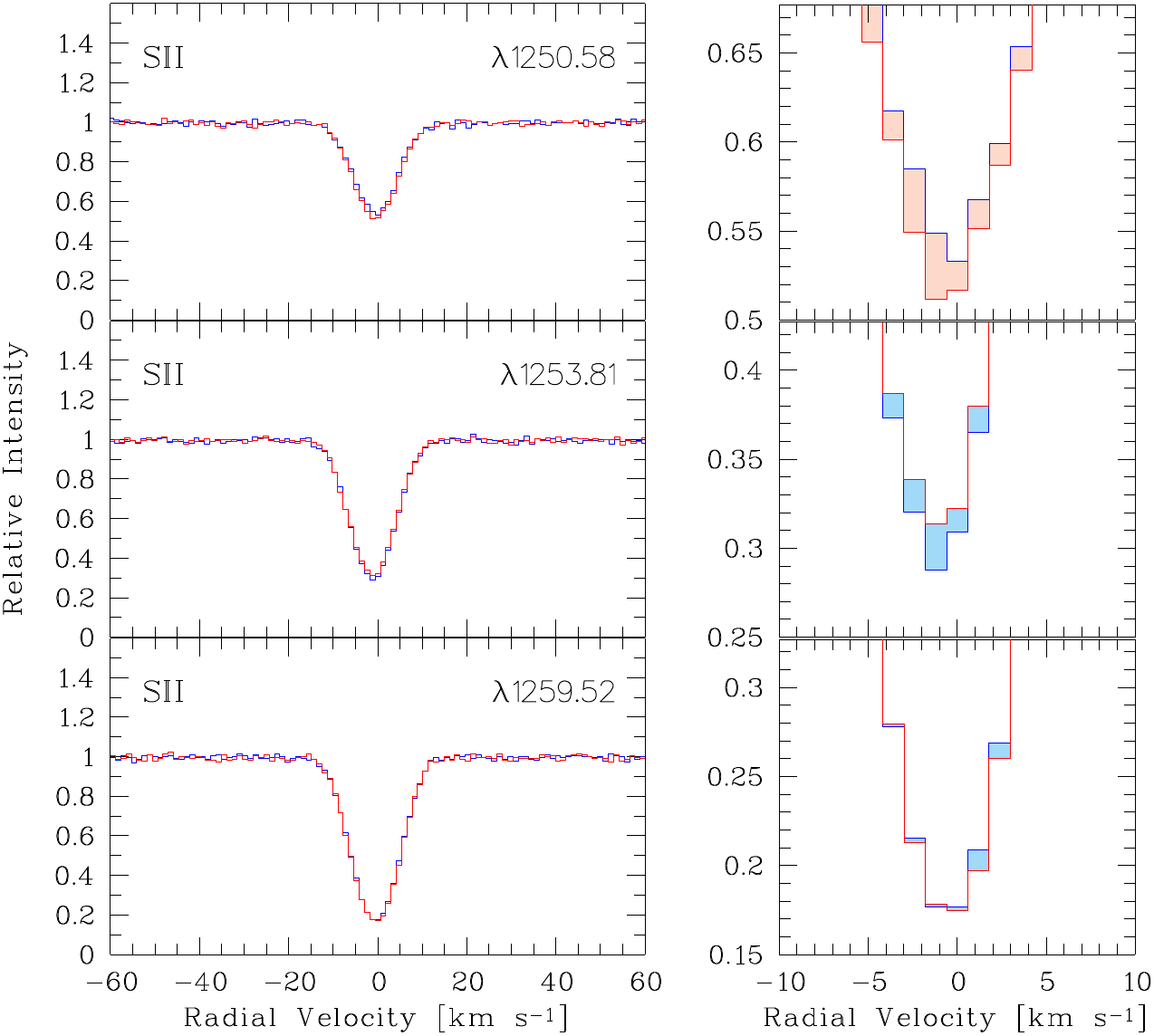}
\caption{Synthetic spectrum (in velocity scale) of a single-component S\,{\sc ii} $\lambda\lambda 1250.58,1253.81,1259.52$ absorption system with a S\,{\sc ii} column density of $N$(S\,{\sc ii}$)=4\times 10^{14}$ cm$^{-2}$, a Doppler parameter of $b=5$ km\,s$^{-1}$, a spectral resolution of $R=45,000$, and a S/N of 100 per resolution element for the source-in-disk scenario. Such a S\,{\sc ii} column density would be expected for a DLA with log $N$(H\,{\sc i}$)=20.48$ and a sulfur abundance of $0.1$ solar assuming solar reference abundances of \citet{Asplund2009}. The geometric condition is $\theta_0=90^{\circ}$. (a) The blue solid lines show the spectrum without the magnetic realignment, the red solid lines show the spectrum when $\theta=90^{\circ}$, $\phi=90^{\circ}$. (b) The red-/blue-shaded areas indicate the excess/deficiency of the absorption due to the magnetic realignment.
}\label{S2ob}
\end{figure*}

Simulations are performed in Fig.~\ref{S2ob} to compare the spectrum profiles of the S\,{\sc ii} triplets for a synthetic absorption spectrum of a DLA with and without the magnetic alignment included. Fig.~\ref{S2ob} was designed up to illustrate the effect in a realistic instrumental set up, with a given typical pixel size and S/N for an optical spectrum taken by an 8m-class telescope. The graphical use of steps instead of curves, which is common in absorption spectroscopy, is intended to visualize the pixel-by-pixel noise variations in the data. The three transitions of singly-ionised Sulfur (S\,{\sc ii}; upper ionisation potential 23.3 eV) at $\lambda\lambda 1250.58,1253.81,1259.52$ represent important tracers for neutral and weakly ionised gas in the local interstellar and intergalactic medium and in distant galaxies (e.g., \citealt{2014ApJ...780...76K,Welsh2012,Richter2001,2014A&A...572A.102F}). Being an $\alpha$ element, singly-ionised Sulfur only has a weak depletion into dust grains (e.g., \citealt{Savage1996}). Thus the interstellar Sulfur abundance is often used as a proxy for the $\alpha$-abundance in the gas. In addition, Sulfur has a relatively low cosmic abundance \citep{Asplund2009} and under typical interstellar conditions (in particular in low-metallicity environments) these lines are not saturated. The three lines are observed in the same wavelength region with identical S/N. The important parameters for simulations are presented in the caption, such as the assumption of S\,{\sc ii} column density, etc. The synthetic spectra were generated using the {\tt FITLYMAN} routine \citep{Fontana1995} implemented in the ESO-MIDAS software package. Atomic data were taken from \citet{Morton2003}. To show the effect clearly, we zoom in the spectrum to the radial velocity range $[-10 km{\cdot}s^{-1}, 10 km{\cdot}s^{-1}]$. The enhancement and reduction of the spectral line profile due to the magnetic realignment change among the triplets. Given the fact that meanwhile optical QSO spectra reach up to a new standard of S/N of a few hundred (e.g., \citealt{D'Odorico16}), the predicted effect is already VISIBLE, if the component structure of the DLA allows a detailed investigation.

The maximum and minimum variations of intensity for different emission lines due to the magnetic realignment are presented in Table \ref{emiretable}. The fluctuation of intensity $\Delta\rm{I}/\rm{I_0}=r^{em}_{mag}-1$. The variations are different among the lines, e.g., for $\theta_0=90^o$, S\,{\sc ii} $\lambda1259.52\mbox{\AA}$ is enhanced up to $21\%$; S\,{\sc i} $\lambda1473.99\mbox{\AA}$ is reduced more than $25\%$ whereas the line S\,{\sc ii} $\lambda1253.81\mbox{\AA}$ is enhanced up to $38\%$. Fe\,{\sc ii} emission spectra in UV and optical band are important for modelling Planetary Nebulae (PNe) \citep{2006ASPC..348..328D} and cloud near AGNs \citep{1985ApJ...288...94W}. The variation of Fe\,{\sc ii}$\lambda1144.93$ emission line due to magnetic realignment are presented in Fig.\ref{Fe2bbremv11to9y6fc90} as an example. In this geometric condition, the magnetic realignment enhances more than $10\%$ of the spectrum when the magnetic field is perpendicular to both the line of sight and the direction of the incident radiation ($\theta=90^{\circ}, \phi=90^{\circ}$).

As demonstrated in Tables 3,4, the modulation due to magnetic realignment is more significant among all the measured multiplets for both absorption and emission spectra.

\begin{table}
\centering
\caption{VARIATIONS OF UV/OPTICAL EMISSION LINES FROM GSA}\label{emiretable}
\begin{tabular}{|c|c|c|c|c|c|}
\hline \hline
Species & $\lambda$ & $(\Delta \rm{I}/\rm{I_0})_{min}$ & $\theta_{0min}$ & $(\Delta \rm{I}/\rm{I_0})_{max}$ & $\theta_{0max}$ \\
 & $(\mbox{\AA})$ & $\%$ & & $\%$ &  \\
\hline
C\,{\sc i} & $1193.03$ & $-20.85$ & $0^{\circ}$ & $+21.00$ & $90^{\circ}$\\
\hline
C\,{\sc ii} & $1334.53$ & $-29.17$ & $0^{\circ}$ & $+28.90$ & $90^{\circ}$\\
\hline
O\,{\sc i} & $1025.76$ & $-14.30$ & $0^{\circ}$ & $+10.44$ & $90^{\circ}$\\
\hline
Si\,{\sc ii} & $1190.42$ & $-27.20$ & $0^{\circ}$ & $+31.49$ & $90^{\circ}$\\
\hline
\multirow{3}{*}{S\,{\sc i}} & $1474.57$ & $-6.32$ & $0^{\circ}$ & $+0.50$ & $54.7^{\circ}$\\
\cline{2-6}
& $1474.38$ & $-19.62$ & $90^{\circ}$ & $+7.49$ & $0^{\circ}$\\
\cline{2-6}
& $1473.99$ & $-38.67$ & $0^{\circ}$ & $+3.89$ & $54.7^{\circ}$\\
\hline
\multirow{2}{*}{S\,{\sc ii}} & $1253.81$ & $-20.51$ & $0^{\circ}$ & $+38.79$ & $90^{\circ}$\\
\cline{2-6}
& $1259.52$ & $-23.04$ & $0^{\circ}$ & $+21.27$ & $90^{\circ}$\\
\hline
S\,{\sc iii} & $1012.50$ & $-32.02$ & $0^{\circ}$ & $+6.01$ & $110^{\circ}$\\
\hline
S\,{\sc iv} & $1062.66$ & $-35.44$ & $0^{\circ}$ & $+44.09$ & $90^{\circ}$\\
\hline
\multirow{3}{*}{Fe\,{\sc ii}} & $1142.36$ & $-2.90$ & $65^{\circ}$ & $+12.10$ & $0^{\circ}$\\
\cline{2-6}
& $1143.22$ & $-9.37$ & $0^{\circ} $& $+22.44$ & $0^{\circ}$\\
\cline{2-6}
& $1144.93$ & $-14.38$ & $0^{\circ}$ & $+13.64$ & $90^{\circ}$\\
\hline
\end{tabular}
\begin{tablenotes}
      \small
      \item Note: Same as Table \ref{absorpretable}, but for resonance emission lines in the UV/optical band.
\end{tablenotes}
\end{table}

\subsection{Submillimeter fine-structure lines}

Submillimeter fine-structure spectra, which arise from magnetic dipole transitions within the ground state, have a broad applicability in astrophysics, such as, determining chemical properties of star-forming galaxies\citep{1999ApJ...514..544K}, predicting the star burst size \citep{2013ApJ...774...68D}, etc. However, previous spectral analysis does not consider the anisotropic radiation and the magnetic realignment. In diffuse ISM and IGM, the influence of GSA on the submillimeter fine-structure absorption lines is exactly the same as that on the UV/optical resonance absorption lines.\footnote{Thanks to such equivalence, the fine-structure absorption lines are not discussed to avoid repetition. Readers may refer to the results in \S\ref{Allowed_tr}.}  The atoms on all the levels in the ground state are magnetically aligned (i.e., only
those matrix tensors $\rho^k_q$ with even k and q = 0 exist) since
the life time of these atoms on all the levels of the ground state are much longer than the magnetic precession
period in diffuse ISM and IGM (see \citealt{YLfine}). For example, the influence of magnetic realignment on C\,{\sc i} $\lambda610{\mu}m$ in a face-on disk is presented in Fig.\ref{C1bbrsubem1c90}. The configuration of the ground state of C\,{\sc i} has 3 levels $3P_{0;1;2}$. C\,{\sc i} $\lambda610{\mu}m$ line represents the transition between the levels within the ground state $3P_{1}$ and $3P_{0}$. Table \ref{submiretable} presents the influence of magnetic fields for a list of submillimeter emission lines.

\begin{table}
\centering
\caption{VARIATIONS OF SUBMILLIMETER LINES DUE TO GSA}\label{submiretable}
\begin{tabular}{|c|c|c|c|c|c|}
\hline \hline
Species & $\lambda$ & $(\Delta \rm{I}/\rm{I_0})_{min}$ & $\theta_{0min}$ & $(\Delta \rm{I}/\rm{I_0})_{max}$ & $\theta_{0max}$ \\
 & $ (\mu m)$ & $\%$ &  & $\%$ &   \\
\hline
[C\,{\sc i}] & $610$ & $-24.2$ & $0^{\circ}$ & $+1.54$ & $54.7^{\circ}$\\
\hline
[C\,{\sc ii}] & $157.7$ & $-9.02$ & $90^{\circ}$ & $+3.68$ & $0^{\circ}$\\
\hline
[O\,{\sc i}] & $63.2$ & $-10.77$ & $0^{\circ}$ & $+0$ & $0^{\circ}$\\
\hline
[Si\,{\sc ii}] & $34.8$ & $-8.76$ & $0^{\circ}$ & $+4.90$ & $90^{\circ}$\\
\hline
[S\,{\sc i}] & $25.2$ & $-12.73$ & $0^{\circ}$ & $+0$ & $0^{\circ}$\\
\hline
[S\,{\sc iii}] & $33.5$ & $-24.67$ & $0^{\circ}$ & $+1.53$ & $54.7^{\circ}$\\
\hline
[S\,{\sc iv}] & $10.5$ & $-2.04$ & $0^{\circ}$ & $+2.00$ & $90^{\circ}$\\
\hline
[Fe\,{\sc ii}] & $26.0$ & $-0.39$ & $35.3^{\circ}$ & $+3.11$ & $90^{\circ}$\\
\hline
\end{tabular}
\begin{tablenotes}
      \small
      \item Note: Same as Table \ref{absorpretable}, but for submillimeter fine-structure lines.
\end{tablenotes}
\end{table}

\subsection{Observations from the medium with turbulent magnetic fields}
\label{turbulence}
\begin{figure*}
\centering
 \subfigure[S\,{\sc i}$\lambda1473.99\mbox{\AA}$ emission]{
\includegraphics[width=0.96\columnwidth]{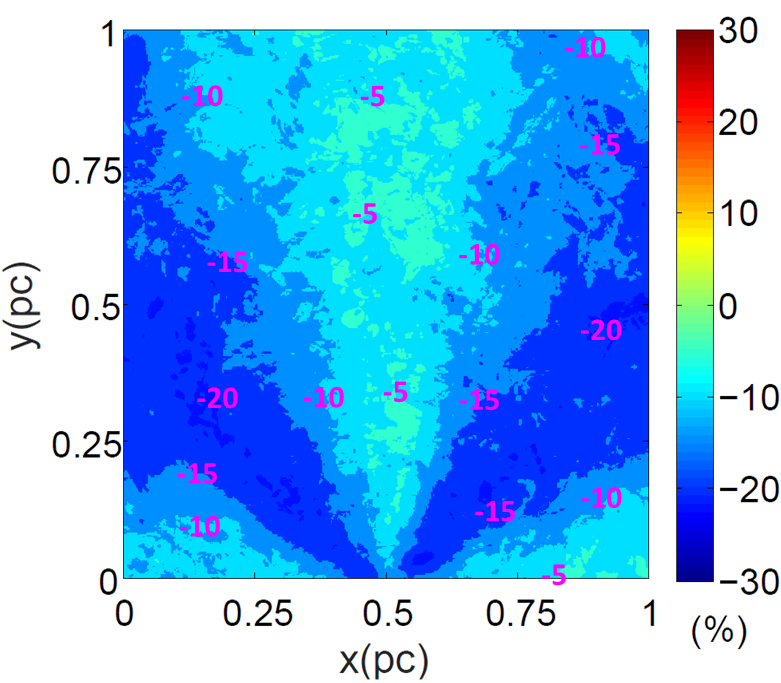}\label{S1uv3to2em_sim}}
 \subfigure[S\,{\sc i}$\lambda1474.38\mbox{\AA}$ emission]{
\includegraphics[width=0.96\columnwidth]{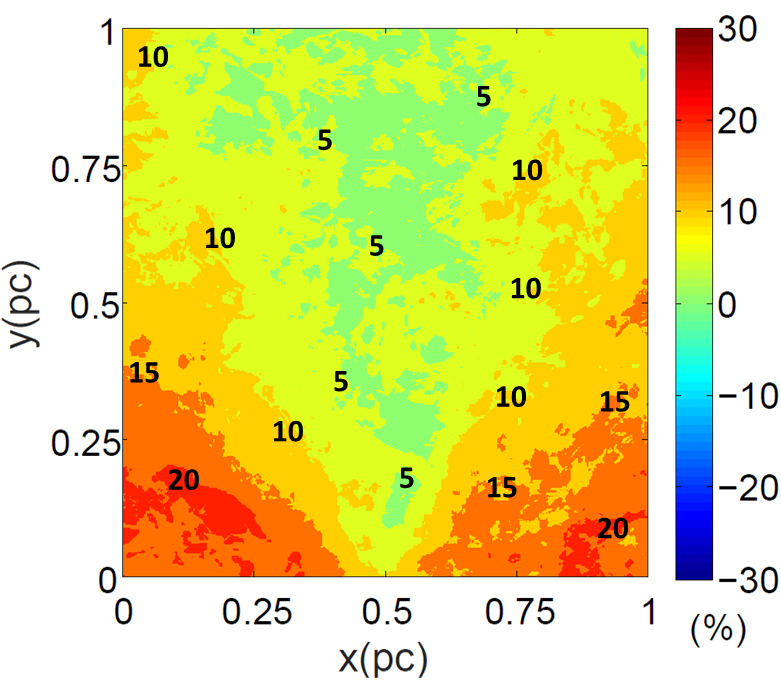}\label{S1uv2to2em_sim}}
\caption{Synthetic observations to super-Alfvenic ISM ($M_a=1.43$) generated by PENCIL MHD code for S\,{\sc i} emission lines (a)$\lambda1473.99\mbox{\AA}$ and (b)$\lambda1474.38\mbox{\AA}$, respectively. The ISM corresponds to an $1pc\times1pc$ area on the picture plane and the optical depth is $0.2pc$. The radiation source, a massive O-type star, resides to the south of the medium $0.1pc$ away. The color scale represents the intensity variation ratio with and without GSA. The percentage of enhancement and depletion are noted with color on the contour.}\label{S1em_sim_cp}
\end{figure*}

In order to address the issue of how much modulation can be induced with the line-of-sight dispersion of magnetic field, we perform synthetic observation on ISM with turbulent magnetic field from numerical simulation. A three-Dimensional (3D) super-Alfvenic ($M_a=1.43$) MHD datacube ($512\times512\times16$), which corresponds to a $1pc(x)\times 1pc(y)\times 0.2pc(z)$ diffuse layer of a reflection nebula, is generated by the MHD-simulation with the PENCIL-code\footnote{See \url{https://code.google.com/archive/p/pencil-code/} for details.}. Note that the simulation is dimensionless and not sensitive to the choice of corresponding physical size. The line-of-sight dispersion is evaluated by the chosen total Alfvenic-Mach Number $M_a$. A massive O-type star radiates UV-photons to illuminate the medium. Synthetic observations are performed on the medium for two emission lines in the S\,{\sc i} triplets $\lambda\sim1474\mbox{\AA}$, which share the same S/N ratio. The intensity variation is integrated along the line of sight and the influence of GSA on these two lines are shown in Fig.~\ref{S1em_sim_cp}. Magnetic alignment induces more than $10\%$ variation and for a sufficient amount of areas more than $20\%$ for both lines. In addition, the variations due to the GSA for the same environment (density, magnetic field, and radiation field ) differ substantially for the two spectral lines. Since the injection scale of interstellar turbulence $\sim100pc$ is much larger than the adopted value here (see \citealt{Armstrong95,CLpwl2010}), the line-of-sight dispersion of the magnetic field in real observations can only be less than that in the synthetic data (Fig.~\ref{S1em_sim_cp}), and thus the intensity variation due to GSA can be more significant in real circumstances.

\section{Influence on physical parameters derived from line ratios}

Many important physical parameters in astronomy are derived from spectral line ratio. For example, the ratio of different transitions of C\,{\sc ii} ($\lambda 1036.34$ and $\lambda 1334.53$) is used to estimate the electron density in different astrophysical environments (see \citealt{2004ApJ...615..767L,2008ApJ...679..460Z,2013ApJ...772..111R} for details). As illustrated in ~\S 3,4, the influence of the GSA varies for different spectral lines. Therefore, the variation of the resulting spectral line ratio is expected to be more significant, e.g., with one line reduced and the other enhanced. According to Eq. \eqref{etai}, the ratio of the intensity between two different absorption lines is given by
\begin{equation}\label{ratioab2}
\begin{split}
\mathcal{R}^{ab}_{\lambda_1,\lambda_2}(\theta_0,\theta_{r},\theta)&=r^{ab}(\lambda_1,\theta_0,\theta_{r},\theta)/r^{ab}(\lambda_2,\theta_0,\theta_{r},\theta),\\
&\propto\frac{\rho^0_0(J'_l)\sqrt{2}+\omega^2_{J'_lJ'_u}\rho^2_0(J'_l)\left(1-1.5\sin^2\theta\right)}{\rho^0_0(J_l)\sqrt{2}+\omega^2_{J_lJ_u}\rho^2_0(J_l)\left(1-1.5\sin^2\theta\right)}.
\end{split}
\end{equation}
And the ratio of the intensities of two emission lines is obtained from Eq. \eqref{epsiloni}:
\begin{equation}\label{ratioem2}
\begin{split}
\mathcal{R}^{em}_{\lambda_1,\lambda_2}(\theta_0,\theta_{r},\theta)&=r^{em}(\lambda_1,\theta_0,\theta_{r},\theta)/r^{em}(\lambda_2,\theta_0,\theta_{r},\theta),\\
&\propto\frac{\rho^0_0(J'_u)+\sum_{\substack{q}}\omega^2_{J'_uJ'_l}\rho^2_q(J'_u)\mathcal{J}^2_{-q}}{\rho^0_0(J_u)+\sum_{\substack{q}}\omega^2_{J_uJ_l}\rho^2_q(J_u)\mathcal{J}^2_{-q}};\\ q&=0, \pm1, \pm2.
\end{split}
\end{equation}
A few examples will be presented in the following subsections.

\subsection{Nucleosynthetic studies in DLAs}

\begin{figure*}
\centering
\subfigure[$\mathcal{R}^{ab}_{1250,2599}$ variation, $\theta_0=90^{\circ}$]{
\includegraphics[width=0.92\columnwidth]{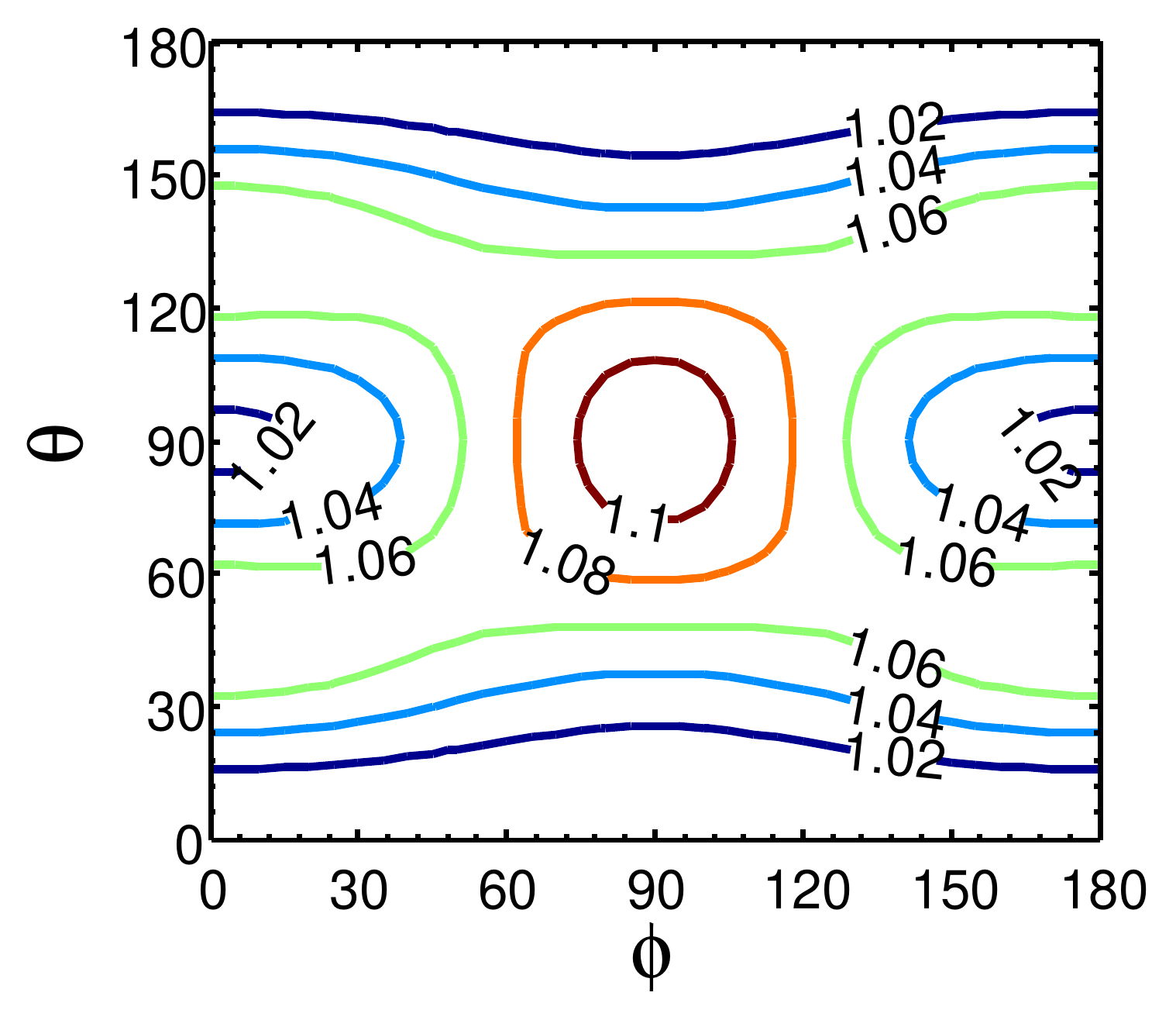}\label{S2Fe2bbrab9to9z6dr1c90}}
\subfigure[$\mathcal{R}^{ab}_{1302,1608}$ variation, $\theta_0=90^{\circ}$]{
\includegraphics[width=0.92\columnwidth]{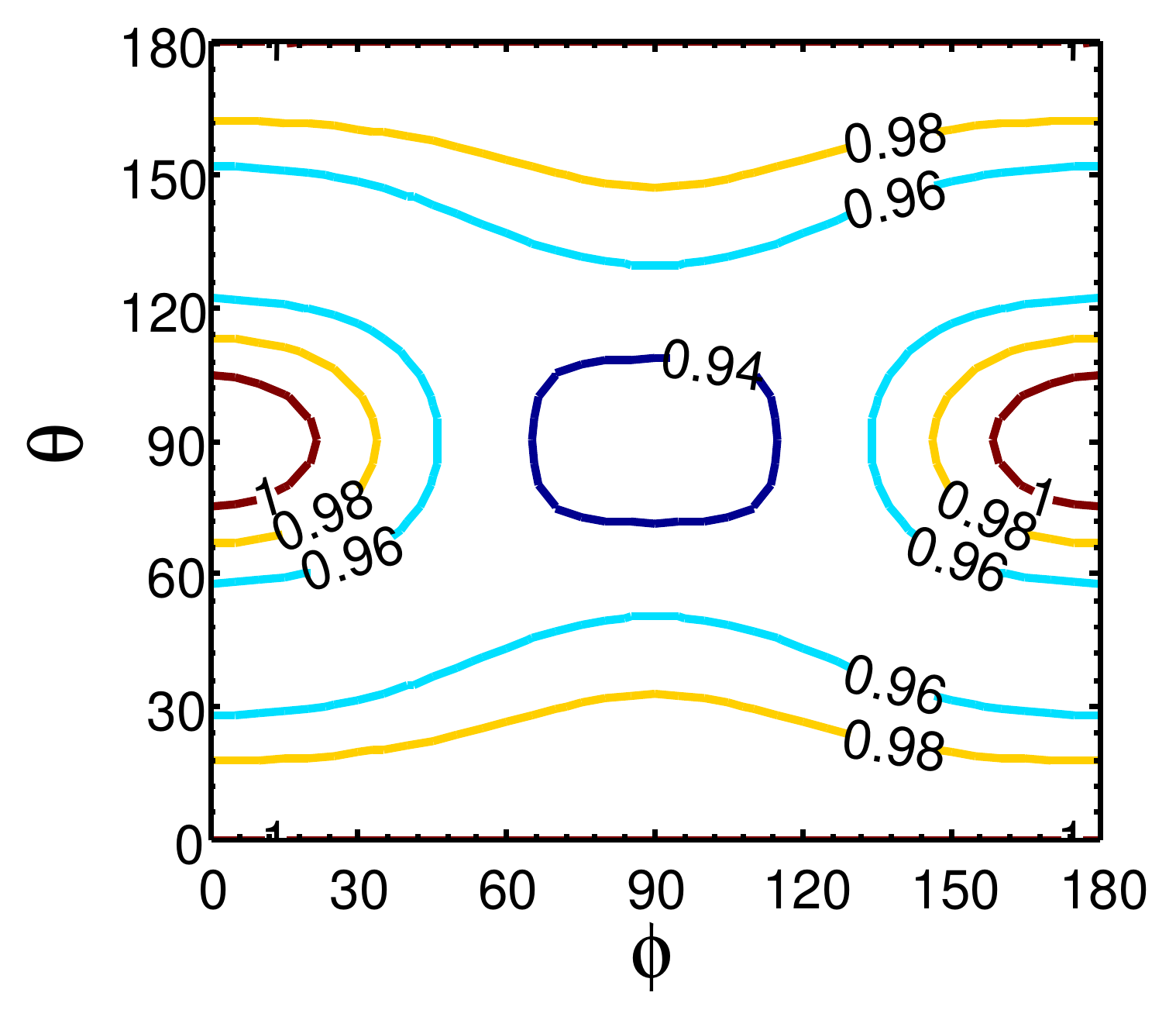}\label{O1Fe2bbrab2to1r9to7c90}}
\caption{Influence of GSA on [$\alpha$/Fe] ratio with (a)$\mathcal{R}^{ab}_{1250,2599}$ (S\,{\sc ii} $\lambda 1250$/Fe\,{\sc ii} $\lambda 2599$) and (b) $\mathcal{R}^{ab}_{1302.17,1608.45}$ (O\,{\sc i} $\lambda 1302.17$/ Fe\,{\sc ii} $\lambda 1608.45$) absorption lines, respectively.}
\end{figure*}

The alpha-to-iron ratio is widely used in spectral analysis. It reflects the nucleosynthetic processes in SFRs (e.g., \citealt{2001ApJS..137...21P}). The $\alpha$ elements include O, Si, S, Ti, etc. Iron refers to Fe peak elements such as Cr, Mn, Co, Fe, etc. $\alpha$ elements in the medium of the galaxy with low metallicity ([Fe/H]$\lesssim$-1.0) are produced exclusively by Type-II supernovae (SNe II), but the [$\alpha$/Fe] ratio suffers a drop when the delayed contribution of Fe from Type-Ia supernovae (SNe Ia) is effective \citep{1997ARA&A..35..503M, 2011MNRAS.417.1534C}. The alpha-to-iron ratio [X/Fe] index (X represents the chemical elements for $\alpha$ elements) is defined by the ratio of abundance observed from the medium in comparison with that from the sun:
\begin{equation}\label{alphafe}
[\rm{X}/\rm{Fe}]\equiv log[\rm{N(X)}/\rm{N(Fe)}]-log[\rm{N(X)}/\rm{N(Fe)}]_{\odot}
\end{equation}
The abundance of the elements is assumed to be equal to the column density of the element in the dominant ionisation state, e.g., Fe\,{\sc ii} for the iron abundance and S\,{\sc ii} for the $\alpha-$element abundance since in DLAs these elements are mostly singly-ionised. Thus, the ratio N(S)/N(Fe) is inferred from the line ratio S\,{\sc ii}/Fe\,{\sc ii}. Nevertheless, the inferred N(S)/N(Fe) is influenced by GSA, as shown in Eq.~\eqref{ratioab2}. The variation of the ratio $\mathcal{R}^{ab}_{\rm{S II, Fe II}}$ with respect to different direction of magnetic fields is presented in Fig.\ref{S2Fe2bbrab9to9z6dr1c90} for the geometric condition $\theta_0=90^\circ$. Furthermore, taking into account of all the possible geometric conditions ($\theta_0$), the maximum and minimum variations for the inferred N(S)/N(Fe) due to GSA is [$-14\%,+10\%$], i.e., [-0.07,+0.04] for [S/Fe].

In comparison, another $\alpha$ element, Oxygen, is presented as an example. Most oxygen in DLAs are neutral and therefore the ratio N(O)/N(Fe) is inferred from O\,{\sc i}/Fe\,{\sc ii}. The influence of GSA on the ratio $\mathcal{R}^{ab}_{\rm{O I, Fe II}}$ is presented in Fig. \ref{O1Fe2bbrab2to1r9to7c90} when $\theta_0=90^\circ$. Comparing Fig.\ref{S2Fe2bbrab9to9z6dr1c90} and Fig.\ref{O1Fe2bbrab2to1r9to7c90}, the inferred N(S)/N(Fe) is enhanced by $10\%$ whereas the inferred N(O)/N(Fe) reduced by $9\%$ when magnetic field is perpendicular to both the line of sight and the direction of the incidental radiation, i.e., [S/Fe] varied by $+0.04$ while [O/Fe] varied by $-0.06$.

\begin{figure}
\centering
\includegraphics[width=0.95\columnwidth]{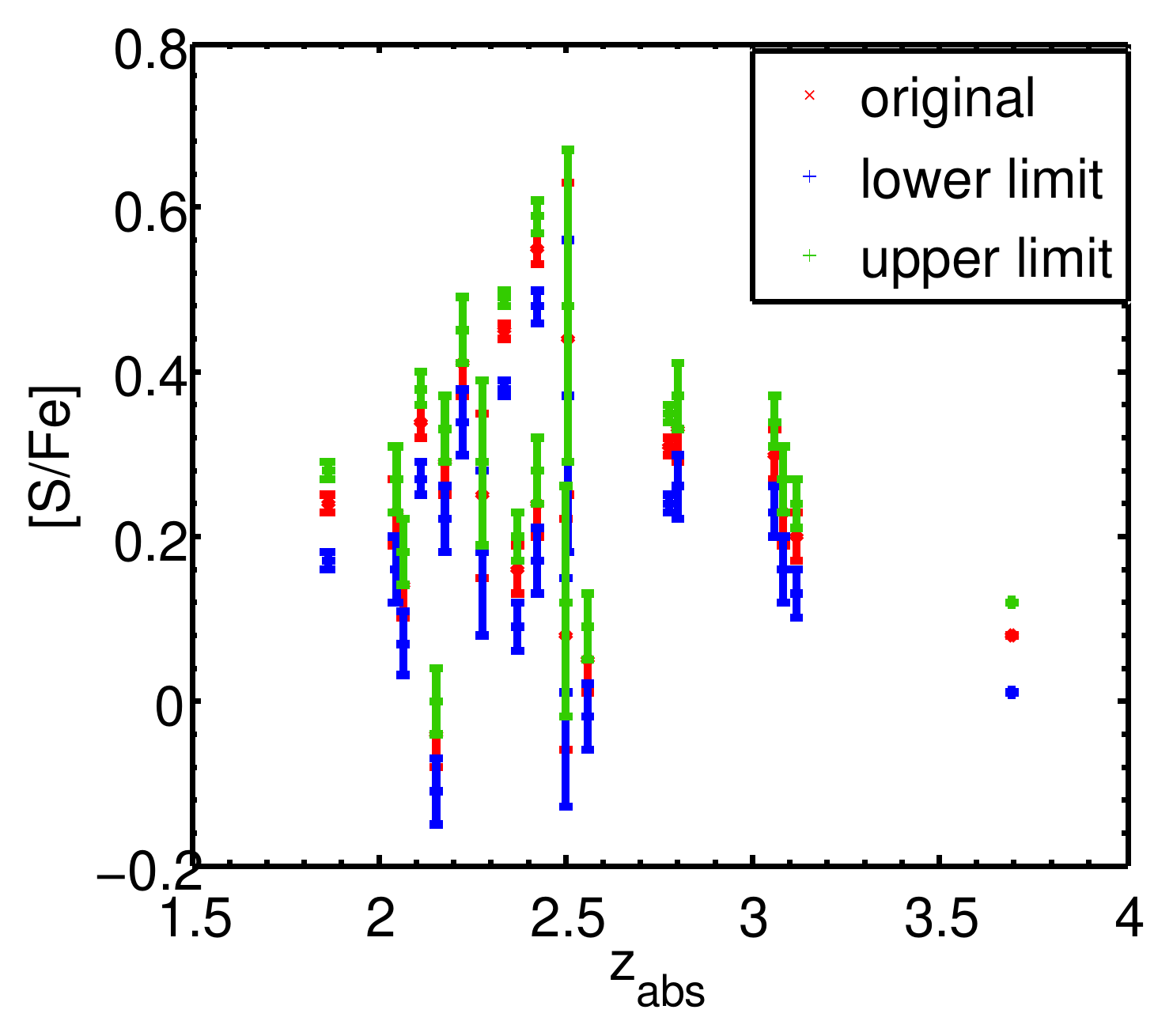}
\caption{Possible influence of GSA on the inferred [S/Fe] from the absorption of DLAs on the QSO spectra. The original data (red points), error bars and the redshift of the corresponding DLA ($z_{abs}$) are obtained from Table 1 in \citet{2008A&A...481..327N}. Green and blue points are the maximum enhancement and reduction of the inferred [S/Fe] when the influence of the magnetic fields is taken into account. The error bars that result from photon noise are independent of the variation due to GSA, and thus are applicable to the upper and lower variation thresholds.}\label{xfecomp}
\end{figure}

Furthermore, Fig.\ref{xfecomp} illustrates the influence of GSA on the [S/Fe] modelled in DLA absorbers on QSO spectra observed by VLT/UVES (see Table 1 in \citealt{2008A&A...481..327N} with X$=$S). By applying the maximum variation of [S/Fe] obtained in this section to the observational data, the upper and lower variation thresholds due to GSA are obtained. As shown in Fig.~\ref{xfecomp}, the variations due to GSA are more significant than the error produced by photon noise. In addition, the depletion of iron in the dust is calculated by [S/Fe] \citep{2008A&A...481..327N}. As demonstrated in Fig.~\ref{xfecomp}, some of the variations due to GSA will even result in the changing sign of [S/Fe].

\subsection{Ionisation studies in diffuse gas}

\begin{figure*}
\centering
\subfigure[$\mathcal{R}^{em}_{1012,1250}$ variation, $\theta_0=90^{\circ}$]{
\includegraphics[width=0.92\columnwidth]{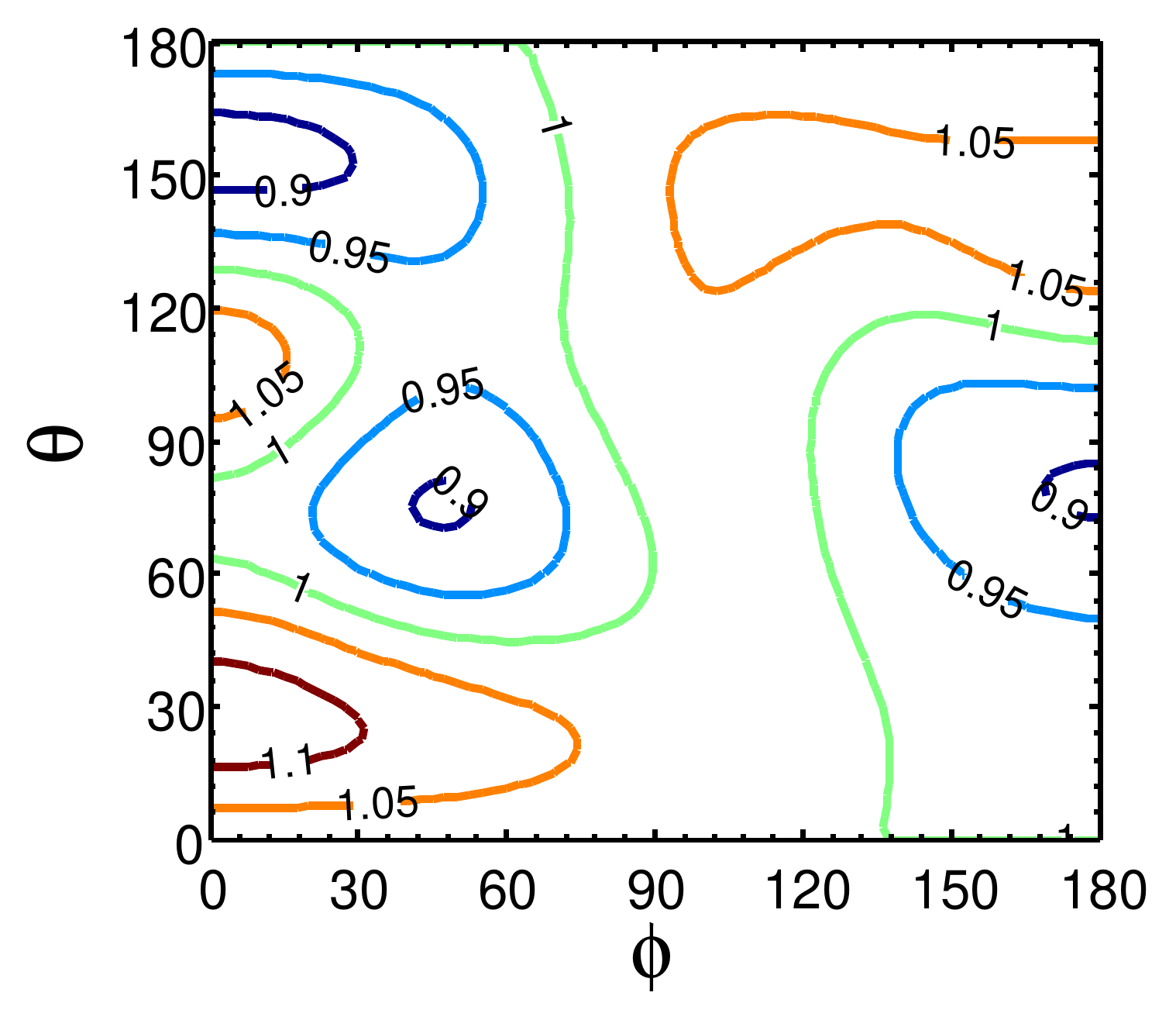}\label{S3S2bbrem1to0pr1c90}}
\subfigure[$\mathcal{R}^{em}_{1012,1250}$ max $\And$ min variations]{
\includegraphics[width=0.92\columnwidth]{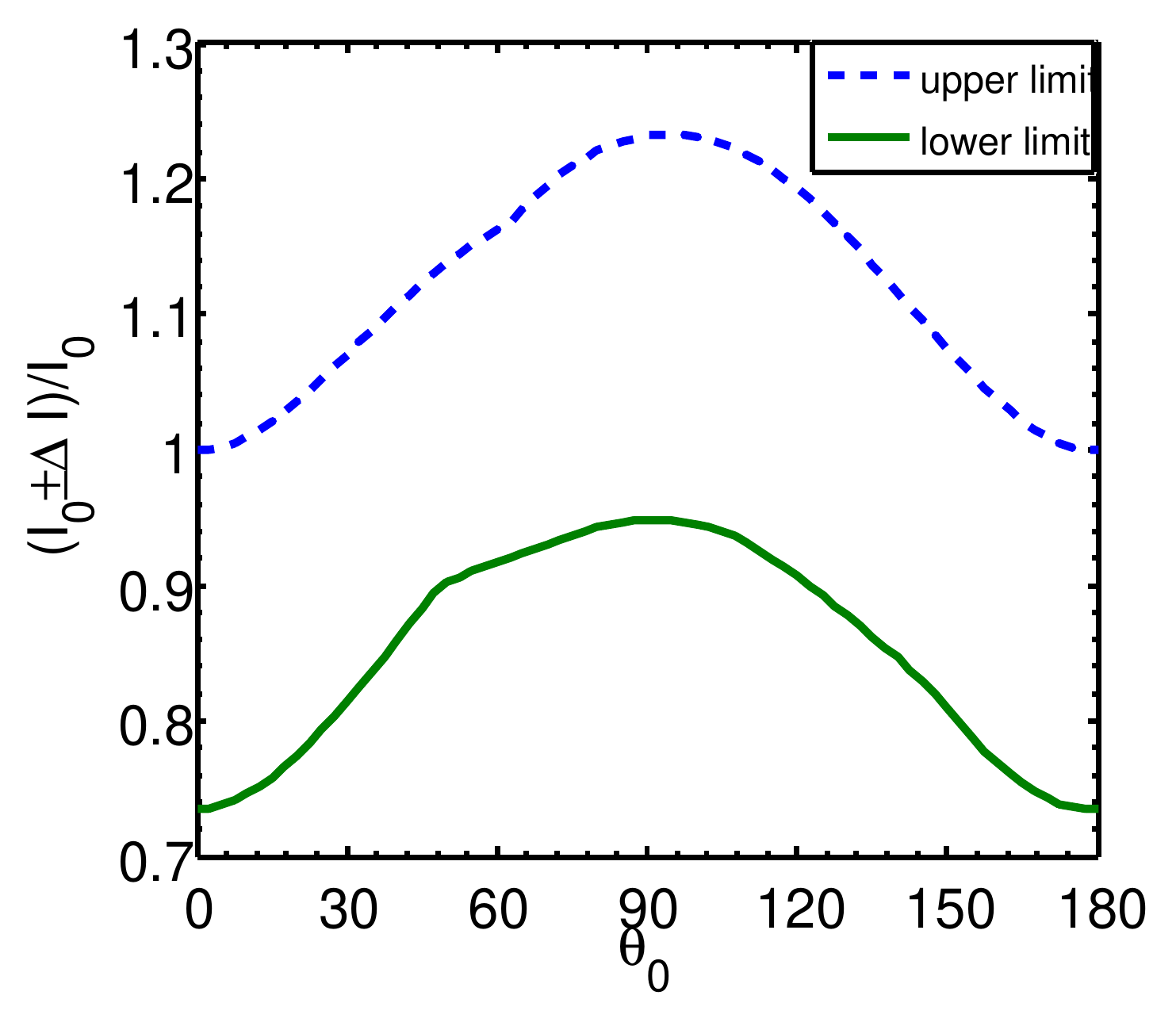}\label{S3S2bbrem1to0pr1}}
\caption{Variation of the ionisation fraction due to GSA: line ratio of S\,{\sc iii} $\lambda1012.49\mbox{\AA}$ and S\,{\sc ii} $\lambda1250.58\mbox{\AA}$ emissions. (a) The variation of line ratio with respect to the direction of the magnetic field in the case of line of sight vertical to the direction of the incident radiation ($\theta_0=90^\circ$); (b) The maximum and minimum variations as a function of $\theta_0$.}
\end{figure*}

The line ratio of the same element from different ionisation states is often used to determine the ionisation fraction (see, e.g., \citealt{2016A&A...590A..68R}). For instance, (S\,{\sc iii})/(S\,{\sc ii}) emission line ratio is employed to determine the ionisation fraction in extragalactic H\,{\sc ii} region (see, e.g., \citealt{1988MNRAS.235..633V}), because higher ionised sulphur is insignificant in most of these H\,{\sc ii} regions \citep{1982ApJ...261..195M, 1985ApJ...291..247M}. The line ratio S\,{\sc iii} $\lambda1012.49$/S\,{\sc ii} $\lambda1250.58$ is adopted to represent (S\,{\sc iii})/(S\,{\sc ii}). Fig.\ref{S3S2bbrem1to0pr1c90} demonstrates the influence of GSA with respect to different magnetic directions for $\theta_0=90^\circ$. The influence of magnetic fields on inferred ionisation fraction varies with different geometric conditions. By taking into account all the possible geometric conditions ($\theta_0$) in Fig.\ref{S3S2bbrem1to0pr1}, the range of variation under the influence of GSA is [$-27\%,+12\%$], which means a variation of [-0.14,+0.05] in log index.

\section{Removing the GSA effect from raw data}

\begin{figure*}
\centering
 \subfigure[Stripe 1 comparison]{
\includegraphics[width=0.99\columnwidth]{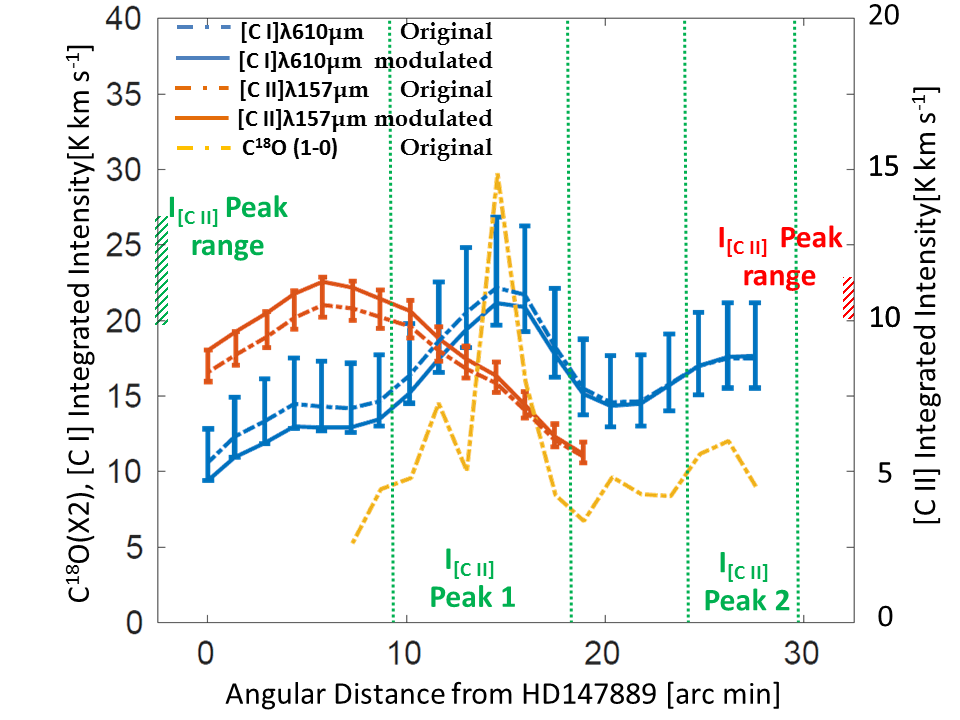}\label{rhoOph1}}
 \subfigure[Stripe 2 comparison]{
\includegraphics[width=0.99\columnwidth]{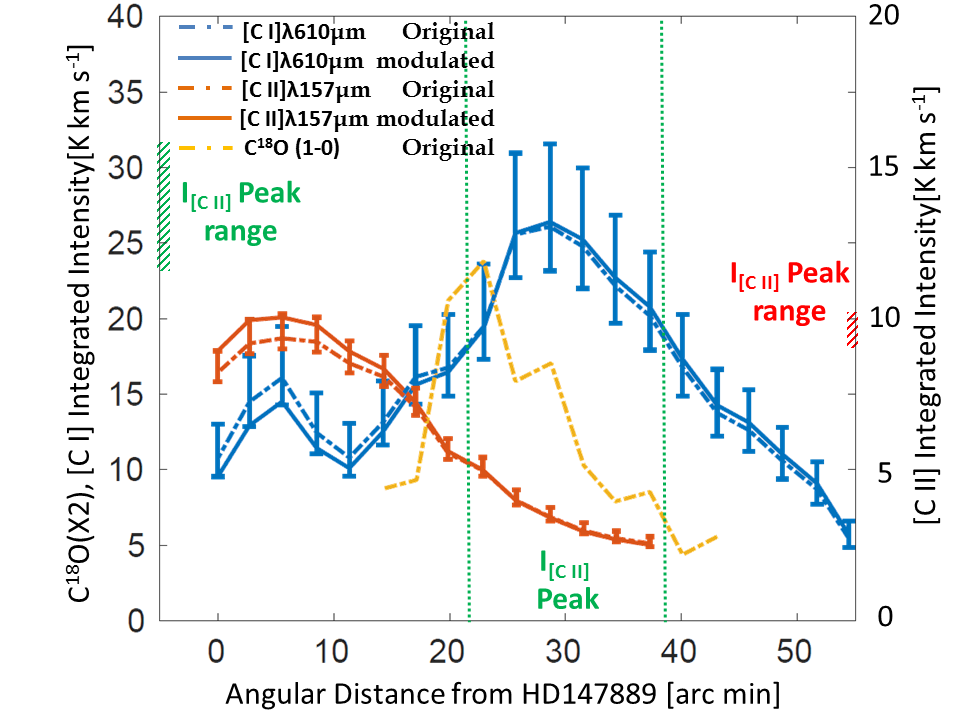}\label{rhoOph2}}
\caption{Modifications of different atomic line intensity due to GSA, the influence of anisotropic pumping and magnetic realignment. Line colors represent: {\color{blue} \bf Blue} for [C\,{\sc i}]$\lambda610{\mu}m$, {\color{RedOrange} \bf Orange} for [C\,{\sc ii}]$\lambda157{\mu}m$, and {\color{YellowOrange} \bf Yellow} for ${\rm C^{18}O(1-0)}$. The raw observational data are in dashed lines from \citet{Kamegai03}. The solid lines are the modifications purely due to anisotropic radiative pumping. The error bars that depict the possible range of the original column density of the corresponding element are produced accounting for the magnetic realignment. Potential peaks for neutral Carbon for both stripes are noted with green dotted lines. The intensity peaks for atomic and single-ionized Carbon lines are marked on the corresponding $y-$axis.
} \label{GSAcali}
\end{figure*}

In this section, we use the observational data of spectral lines from PDRs in $\rho$ Ophiuchi cloud as an example to demonstrate how to account for GSA effect. The cloud, $150pc$ from the earth, is illuminated by the UV/optical radiation from the B\,{\sc iii}V star HD147889, which resides approximately $d_{Source}=0.75pc$ behind the cloud \citep{Liseau99}. Atomic and single-ionized Carbon traces the neutral gas of the PDRs in the cloud whereas molecular lines are also detected (e.g., ${\rm C^{18}O}$ \citealt{Kamegai03}). Two stripes in the cloud are used here as examples to perform the analysis. Dashed lines plotted in Fig.~\ref{GSAcali} are adopted from the observational analysis in \citet{Kamegai03}. We first consider only the radiative alignment. We adopt the fitting curves for [C\,{\sc i}] and [C\,{\sc ii}] lines from Table~\ref{pumpfit} in \S 3:
\begin{equation}\label{fitC1C2}
\begin{split}
r^{pump}_{[C\,{\sc i}]}(\theta_0)&=1.0332+0.094\cos2\theta_0,\\
r^{pump}_{[C\,{\sc ii}]}(\theta_0)&=0.9795-0.0599\cos2\theta_0,\\
\theta_0=-\arctan\frac{r}{d_{Source}}&=-\arctan\left[5.82\times10^{-2}(r_{\theta_0}/arc min)\right],\\
\end{split}
\end{equation}
where $r_{\theta_0}$ is the picture plane angular distance from the source to the analyzed medium in $arc min$. The intensity ($I_{VV}$) after correcting the effect from radiative alignment is thus obtained from the observational intensity ($I_{ob}$) by:
\begin{equation}\label{Caliori}
I_{VV}=I_{ob}/r^{pump}(\theta_0).
\end{equation}
The results are shown with solid lines in Fig.~\ref{GSAcali}. The Van Vleck angle flipping is seen at $\sim24\arcmin$. We further take into account the magnetic realignment. The picture plane magnetic field in $\rho$ Ophiuchi cloud is obtained through star polarization in \citet{Kwon15}. The unknown component of magnetic field along the line of sight leads to some uncertainties in the correction, as marked by the error bars in Fig.~\ref{GSAcali}. The potential neutral Carbon peak can be shifted from the original one in raw data (dashed lines). The intensity of atomic and single-ionized Carbon can be significantly modified by GSA. The real column density distribution of neutral Carbon may be synchronized with or largely deviated from that of ${\rm C^{18}O}$. Therefore, more accurate magnetic field analysis with compatible resolution to the spectral analysis has to be performed in such area for a proper study of the interstellar gas.

\section{Discussion}

\begin{figure*}
\centering
\subfigure[]{
\includegraphics[width=0.92\columnwidth]{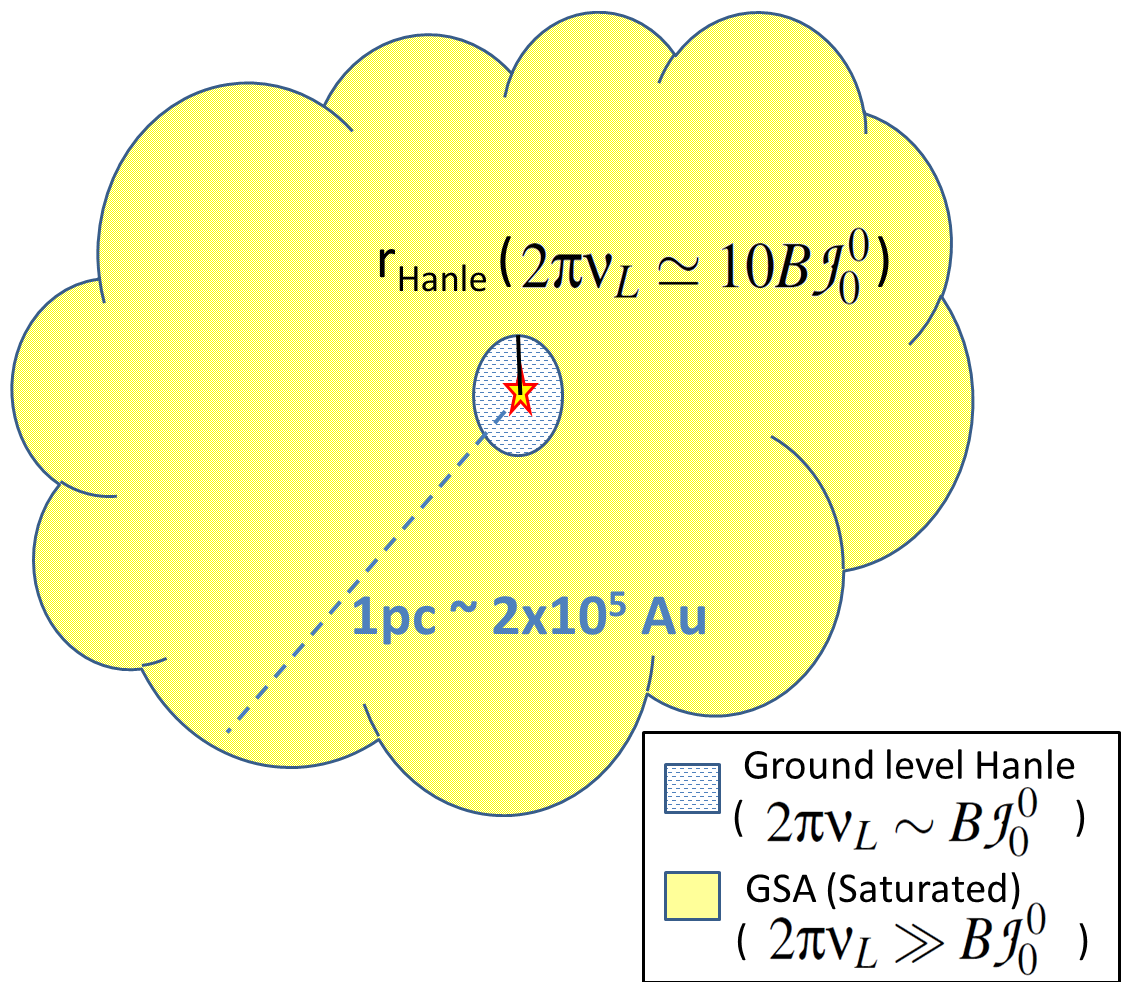}\label{RegimeGeo}}
\subfigure[Ground level Hanle effect Radius]{
\includegraphics[width=0.92\columnwidth]{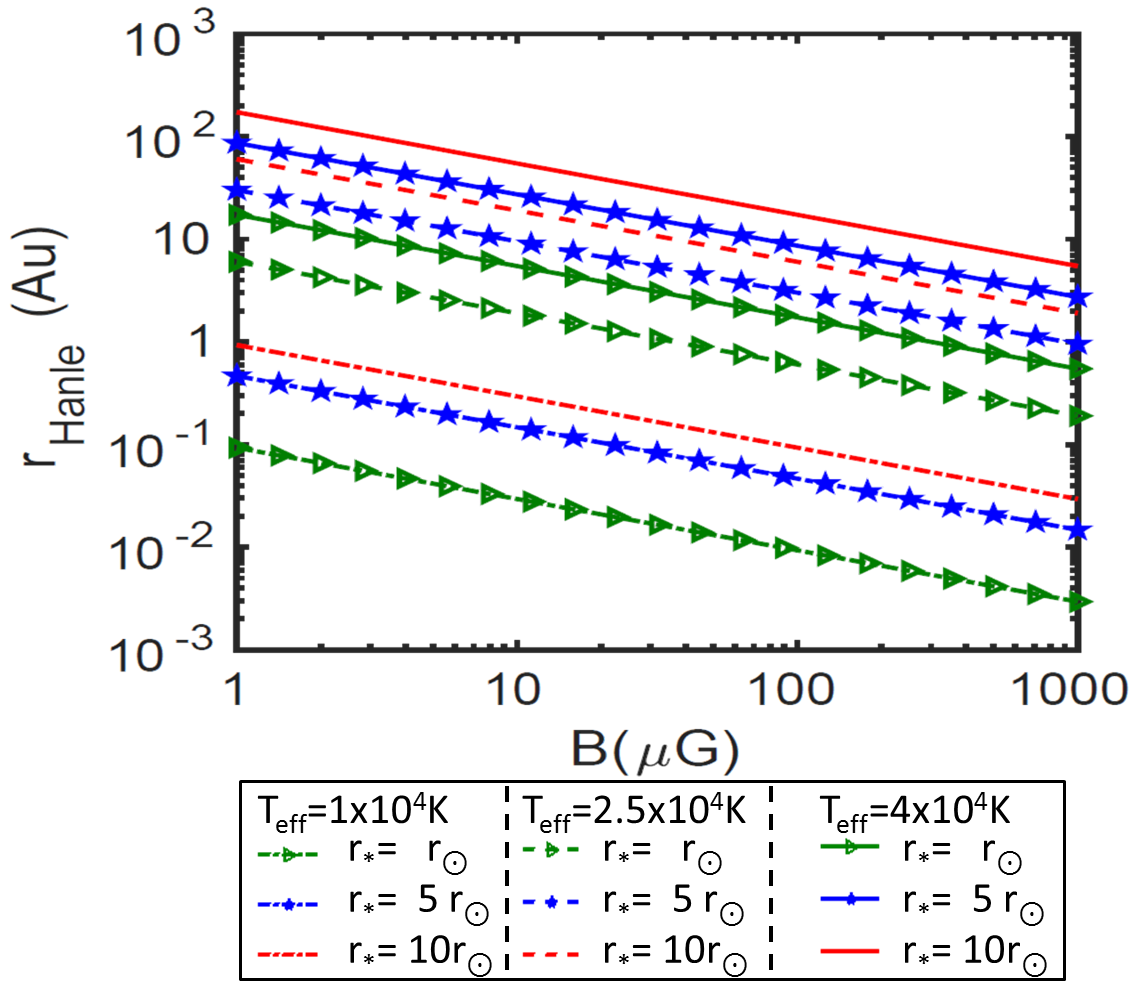}\label{RegimeRadius}}
\caption{(a)Geometric illustration of different regimes with the corresponding scale; The circle denotes the boundary between ground-level Hanle regime and the GSA regime in this paper; (b) The boundary scale of ground-level Hanle effect with different physical environments: star radius $r_\ast$, effective temperature of the pumping source $T_{eff}$, and the magnetic field strength ($x-$axis). $r_\odot$ is the solar radius.}
\end{figure*}

As illustrated in this paper, the variation of spectral line intensity induced by GSA varies among different spectral lines. Thus, such influence could be precisely analysed if multi-spectral lines for the same element is achievable. In addition, the alignment on the ground state is transferred to the levels of atoms on the excited states through absorption process, as illustrated in \S 3 and 4. The intensity of the ultraviolet-pumped fluorescence lines, which are derived from successive decays to different levels of atoms and applied to the modelling of reflection nebulae \citep{1984ApJ...277..623S,1986ApJ...305..399S}, are dependent on the initial upper levels, and thus is influenced by GSA through the scattering process. The influence of collision is neglected in this paper, which applies to most diffuse ISM and IGM. Collisions reduce the alignment efficiency (see \citealt{Hawkins:1955fv}). The collision effect can become important in the case of higher density medium where the collision rate $\tau_c^{-1}$ (either inelastic collision rate or Van der Waals collision rate) dominates over optical pumping rate $B_{lu}\bar{J}^0_0$ (see \citealt{YLfine} for details). The focus of our paper is on the GSA effect that is a saturated state. When the magnetic precession rate is comparable to the optical pumping from the ground state ($2\pi\nu_L\sim B_{lu}\bar{J}^0_0$), the ground-level Hanle effect is applicable (see \citealt{Landolfi:1986lh}), which means the magnetic field influence on the spectrum is not limited to the change of direction but also the magnetic field strength. As demonstrated in \citet{YLHanle}, the effect becomes saturated when $2\pi\nu_L\rightarrow 10B_{lu}\bar{J}^0_0$. Therefore, we set $r_{Hanle}$where $2\pi\nu_L\simeq 10B_{lu}\bar{J}^0_0$ as the boundary between the ground-level Hanle regime and the GSA regime (see Fig.~\ref{RegimeGeo}):
\begin{equation}\label{rHanle}
r_{Hanle}=r_\ast\sqrt{\frac{A_{ul}[J_u]}{1.76(B/{\mu}G)(\exp(h\nu/(k_BT))-1)[J_l]}}
\end{equation}
We calculate the boundary $r_{Hanle}$ for C\,{\sc ii}$\lambda1037\mbox{\AA}$ in the presence of stars with different effective temperature $T_{eff}$ and radius $r_\ast$ for the magnetic field with the strength range from ${\mu}G$ to $mG$ in Fig.~\ref{RegimeRadius}. The $r_{Hanle}$ increases as the magnetic field becomes weaker and as the effective temperature $T_{eff}$ and radius $r_\ast$ increase. Nevertheless, even in the most optimistic scenario with $T_{eff}=4\times10^4K, r_\ast=10r_\odot, B=1{\mu}G$ (though rarely applies), the boundary $r_{Hanle}\simeq180Au=8.5\times10^{-4}pc$, which is a thousand times smaller than the normal H\,{\sc ii} Region which is in $pc$ scale. All the analysis with GSA and their observational implication in this paper is applicable to most of the ISM, except when performing very high resolution spectral analysis on regions very close to the very bright O-type star.

Indeed, many spectral lines we measured reside in multiplets. It is worth noting that the influence of GSA becomes more significant for multiplets and line ratios. Furthermore, our work reveals that impact of magnetic fields on the analysis of interstellar environment cannot be neglected. The study is complimentary to the interstellar magnetic field diagnoses.

\section{Conclusions}

We have demonstrated the influence of GSA effect on the spectroscopy. We emphasize that GSA is a general physical process that induces visible systematic variations to both absorption and emission spectral lines observed from diffuse medium, e.g., DLAs, H\,{\sc ii} Regions, PDRs, SFRs, Herbig Ae/Be disks, etc. Comprehensive results are provided to demonstrate the influence of the GSA effect -including radiative alignment and magnetic realignment- on resonance UV/optical absorption/emission lines and submillimeter fine-structure lines observed from different astrophysical environments. Synthetic observations are performed to present such influence on spectral line profile and to investigate the influence of GSA in turbulent magnetic fields. Variations of the physical parameters inferred from line ratios due to GSA are studied. We illustrate how to remove the GSA effect from raw data. Our main conclusions are:

\begin{itemize}
\item Measurable modulations are induced on the absorption and emission lines observed from diffuse ISM and IGM due to GSA.
\item The influence of GSA on the spectral line intensity is not diminished due to line-of-sight dispersion of turbulent magnetic fields such as in ISM and IGM.
\item The enhancement and reduction of the same spectrum line change in accordance with the direction of the magnetic field and the radiation geometry.
\item The variation of the line intensity changes from line to line. As a result, the influences on the multiplets and line ratios are even more distinct, affecting the inferred physical parameters.
\item The analytical model set up in this paper can be used for correcting the  GSA effect from interstellar spectroscopy. Without such correction, the physical environment inferred can not achieve the accuracy that current instruments promise.
\item GSA should be considered in the future spectral analysis.
\end{itemize}

\section*{Acknowledgements}
We are grateful to Gesa Bertrang, Reinaldo Santos de Lima, Ruoyu Liu, Michael Vorster for the helpful discussions. We thank the referee for valuable comments and suggestions.

%




\bibliographystyle{mnras}
\bibliography{Zhan0306} 




\appendix

\section{BASIC FORMULAE ON ATOMIC ALIGNMENT}\label{formula}

In this Appendix, we illustrate the basic equations on atomic alignment. The calculations in this Appendix are all performed in the theoretical frame $xyz-$system in Fig.1(a).

Anisotropic radiation excites atoms through photo-excitation and consequently results in spontaneous emissions. Occupations of the atoms on different levels of the ground state will alter when there exists anisotropic radiation field. Magnetic realignment will redistribute the angular momenta of the atoms due to fast magnetic precession, which only happens on the ground state in general ISM and IGM (see \citealt{2012JQSRT.113.1409Y,2015ASSL..407...89Y} for details). The equations to describe the evolution on upper and lower levels are \citep[see][]{Landolfi:1986lh,landi2004}:
\begin{equation}\label{upperevol}
\begin{split}
&\dot{\rho}_q^k(J_u)+2\pi i\nu_L g_u q\rho_q^k(J_u)=-\sum_{\substack{J_l}}A(J_u\rightarrow J_l)\rho_q^k(J_u)+\sum_{\substack{J_{l}k'}}[J_l]\\
&\times\left[\delta_{kk'}p_{k'}(J_u,J_l)B_{lu}\bar{J}^0_0+\sum_{\substack{Qq'}}r_{kk'}(J_u,J_l,Q,q')B_{lu}\bar{J}^2_Q\right]\rho^{k'}_{-q'}(J_l),
\end{split}
\end{equation}
\begin{equation}\label{groundevol}
\begin{split}
\dot{\rho}_q^k(J_l)&+2\pi i\nu_L g_l q\rho_q^k(J_l)=\sum_{\substack{J_u}}p_k(J_u,J_l)[J_u]A(J_u\rightarrow J_l)\rho_q^k(J_u)\\
&-\sum_{\substack{J_{u}k'}}\left[\delta_{kk'}B_{lu}\bar{J}^0_0+\sum_{\substack{Qq'}}s_{kk'}(J_u,J_l,Q,q')B_{lu}\bar{J}^2_Q\right]\rho^{k'}_{-q'}(J_l),
\end{split}
\end{equation}
in which
\begin{equation}
\begin{split}
&p_k(J_u,J_l)=(-1)^{J_u+J_l+1}\left\{\begin{array}{ccc}J_l&J_l&k\\J_u&J_u&1\end{array}\right\},\\
&p_0(J_u,J_l)=\frac{1}{\sqrt{[J_u,J_l]}},\\
&r_{kk'}(J_u,J_l,Q,q)=(3[k,k',2])^{\frac{1}{2}}\left\{\begin{array}{ccc}1&J_u&J_l\\1&J_u&J_l\\2&k&k'\end{array}\right\}\left(\begin{array}{ccc}k&k'&K\\q&q'&Q\end{array}\right),\\
&s_{kk'}(J_u,J_l,Q,q)=(-1)^{J_l-J_u+1}[J_l](3[k,k',K])^\frac{1}{2}\\
&\times\left(\begin{array}{ccc}k&k'&2\\q&q'&Q\end{array}\right) \left\{\begin{array}{ccc}1&1&2\\J_l&J_l&J_u\end{array}\right\}\left\{\begin{array}{ccc}k&k'&2\\J_l&J_l&J_l\end{array}\right\}.
\end{split}
\end{equation}
The quantities $J_{u}$ and $J_{l}$ are the total angular momentum quantum numbers for the upper and lower levels, respectively. The quantities $\rho_q^k$ and $\bar{J}_Q^K$ are irreducible density matrices for the atoms and the incident radiation, respectively. $6-j$ and $9-j$ symbols are represented by the matrices with $"\{$ $\}"$, whereas $3-j$ symbols are indicated by the matrices with $"()"$ (see \citealt{1989PhT....42l..68Z} for details). {\em The second terms on the left side of Eq.~\eqref{upperevol} and Eq.~\eqref{groundevol} stand for the magnetic realignment.} The two terms on the right side represent spontaneous emissions and the excitations from lower levels. Note that the symmetric processes of spontaneous emission and magnetic realignment conserve $k$ and $q$. Therefore, the steady state occupations of atoms on the ground state are obtained by setting the left side of Eq.~\eqref{upperevol} and Eq.~\eqref{groundevol} zero\footnote{This is the correct version of Eq. (8) in \citet{YLHanle}, in which there is a typo. The term $\Gamma$ in Eq. (8) of \citet{YLHanle} denotes $B_{lu}$ as a common factor. But that will be an error for multi-level atoms since the Einstein coefficients for transitions between different upper and lower levels are different.}:
\begin{equation}\label{groundoccu}
\begin{split}
&2\pi i\rho_q^k(J_l)qg_l\nu_L-\sum_{\substack{J_uk'}}\bigg\{p_k(J_u,J_l)\frac{[J_u]}{\sum_{J''_{l}}A''/A+i\Gamma'q}\\ &\times\sum_{\substack{J'_l}}B_{lu}[J'_l]\left[\delta_{kk'}p_{k'}(J_u,J'_l)\bar{J}^0_0+\sum_{\substack{Qq'}}r_{kk'}(J_u,J'_l,Q,q')\bar{J}^2_Q\right]\\
&-\delta_{J_lJ'_l}\left[\delta_{kk'}B_{lu}\bar{J}^0_0+\sum_{\substack{Qq'}}B_{lu}s_{kk'}(J_u,J_l,Q,q')\bar{J}^2_Q\right]\bigg\}\rho^{k'}_{-q'}(J'_l)=0
\end{split}
\end{equation}
where $\Gamma'$ equals $2\pi\nu_L g_u/A$. Magnetic realignment on the levels in excited states can be neglected because the spontaneous emission rate from the excited states is much higher than magnetic precession rate in diffuse ISM and IGM. As a result, $\Gamma'\simeq0$. On the other hand, the magnetic precession rate is much higher than the photon excitation rate of the atoms on the ground state in the diffuse media of ISM and IGM ($\nu_L\gg B_{lu}\bar{J}^0_0$). Thus, Eq.~\eqref{groundoccu} is making sense only when $q=0$ so that the first term on the left equals 0. By solving the above equations, the atomic density tensors on different levels ($\rho^k_q(J_l),\rho^k_q(J_u)$) under the influence of atomic alignment are obtained.


\bsp	
\label{lastpage}
\end{document}